\documentclass[aps,pra,twocolumn,showpacs,amsmath,amssymb,floatfix]{revtex4-1}

\usepackage{graphicx}
\usepackage{amsmath}
\usepackage{epsfig, float}
\usepackage{natbib}
\usepackage{bm}
\usepackage{latexsym}

\newcommand{\ketPsi}[1]{\left| \Psi(#1) \right>}
\newcommand{\om}[2]{\omega_{#1}(#2)}

\newcommand{\tinveffm}[4]{\left[ \frac{1}{m^{\ast}_{#3}(#4)} \right]^{#1 #2}}

\newcommand{\ketBloch}[2]{\left| \psi_{#1 #2} \right>}
\newcommand{\ketModBloch}[2]{\left| \phi_{#1 #2} \right>}
\newcommand{\expvalS}[2]{\left< \Psi(#2) \middle| #1 \middle| \Psi(#2) \right>}
\newcommand{\rspaceket}[1]{\left< \mathbf{r} \middle| #1 \right>}
\newcommand{\rspaceketBloch}[2]{\left< \mathbf{r} \middle| \psi_{#1 #2} \right>}
\newcommand{\modmom}[4]{\mathfrak{P}^{#1}_{#2 #3}(#4)}
\newcommand{\bLax}[3]{\boldsymbol \xi_{#1 #2} (#3)}
\newcommand{\Lax}[4]{\xi^{#1}_{#2 #3} (#4)}
\newcommand{\meBloch}[5]{\left< \psi_{#2 #3} \middle| #1 \middle| \psi_{#4 #5} \right>}
\newcommand{\meUBloch}[5]{\left< u_{#2 #3} \middle| #1 \middle| u_{#4 #5} \right>}
\newcommand{\posOp}[1]{\hat{r}^{#1}}
\newcommand{\momOp}[1]{\hat{p}^{#1}}
\newcommand{\mommat}[4]{p^{#1}_{#2 #3}(#4)}
\newcommand{\ketWvpMB}[2]{\left| \bar{\phi}_{#1}(#2) \right>}
\newcommand{\expvalWvpMB}[3]{\left< \bar{\phi}_{#2}(#3) \middle| #1 \middle| \bar{\phi}_{#2}(#3) \right>}

\newcommand{\gphase}[3]{\gamma_{#1}(#2, #3)}
\newcommand{\modforce}[4]{\mathfrak{F}^{#1}_{#2 #3}(#4)}
\newcommand{\vg}[3]{v^{\text{g},#1}_{#2}(#3)}
\newcommand{\van}[3]{v^{\text{an},#1}_{#2}(#3)}
\newcommand{\vosc}[4]{\mathcal{V}^{#1}_{#2}(#3, #4)}
\newcommand{\bvoscint}[1]{\left< \mathbf{v}^{\text{osc}}(#1) \right>}
\newcommand{\voscintpar}[1]{\langle v^{\text{osc}}_{\parallel}(#1) \rangle}
\newcommand{\voscintperp}[1]{\left< v^{\text{osc}}_{\perp}(#1) \right>}
\newcommand{\bvanint}[1]{\left< \mathbf{v}^{\text{an}}(#1) \right>}
\newcommand{\bveffint}[1]{\left< \mathbf{v}^{\text{g}}(#1) \right>}

\newcommand{\tv}[5]{\mathfrak{J}^{#1 #2}_{#3 #4}(#5)}
\newcommand{\tosc}[5]{\mathcal{J}^{#1 #2}_{#3}(#4,#5)}

\newcommand{\ataosc}[5]{\tilde{\mathcal{K}}^{#1 #2}_{#3}(#4,#5)}
\newcommand{\staosc}[5]{\bar{\mathcal{K}}^{#1 #2}_{#3}(#4,#5)}
\newcommand{\bbryc}[2]{\boldsymbol \Omega_{#1}(#2)}
\newcommand{\bryc}[3]{\Omega^{#1}_{#2}(#3)}
\newcommand{\bbrycosc}[3]{\boldsymbol \Lambda_{#1}(#2,#3)}
\newcommand{\brycosc}[4]{\Lambda^{#1}_{#2}(#3,#4)}
\newcommand{\bbrycwvp}[2]{\boldsymbol \Omega_{#1}(#2)}
\newcommand{\brycwvp}[3]{\Omega^{#1}_{#2}(#3)}
\newcommand{\atosc}[5]{\tilde{\mathcal{J}}^{#1 #2}_{#3}(#4,#5)}
\newcommand{\stosc}[5]{\bar{\mathcal{J}}^{#1 #2}_{#3}(#4,#5)}
\newcommand{\LaxSq}[7]{X^{#1,#2 #3}_{#4 #5}(#6, #7)}
\newcommand{\axvaosc}[4]{\Xi^{#1}_{#2}(#3,#4)}

\newcommand{\pder}[1]{\frac{\partial}{\partial #1}}
\newcommand{\pdder}[2]{\frac{\partial^2}{\partial #1 \partial #2}}
\newcommand{\expval}[1]{\left< #1 \right>}
\newcommand{\comm}[2]{\left[ #1 , #2 \right]}
\newcommand{\ket}[1]{\left| #1 \right>}
\newcommand{\levic}[3]{\epsilon^{#1 #2 #3}}

\newcommand{\refeq}[1]{Eq.~\eqref{#1}}
\newcommand{\refeqs}[1]{Eqs.~\eqref{#1}}
\newcommand{\refsec}[1]{Sec.~\ref{#1}}
\newcommand{\refsecs}[1]{Secs.~\ref{#1}}
\newcommand{\reffig}[1]{Fig.~\ref{#1}}
\newcommand{\reffigs}[1]{Figs.~\ref{#1}}

\newcommand{\mbf}{\mathbf}

\newcommand{\bposOp}{\hat{\mathbf{r}}}
\newcommand{\bmomOp}{\hat{\mathbf{p}}}
\newcommand{\bpos}{\mathbf{r}}
\newcommand{\pos}{r}

\newcommand{\baccOp}{\hat{\mathbf{a}}}
\newcommand{\velOp}{\hat{v}}

\newcommand{\bvelOp}{\hat{\mathbf{v}}}
\newcommand{\blattvec}{\mathbf{R}}
\newcommand{\HOp}{\hat{H}}
\newcommand{\intBZ}{\int_{\text{BZ}} d \mathbf k \,}
\newcommand{\intallpos}{\int d \mathbf{r} \, }
\newcommand{\intunitcell}{\int_{V_\text{cell}} \frac{d \mathbf{r}}{V_\text{cell}} \, }
\newcommand{\foMix}{\Delta}
\newcommand{\bkmov}{\boldsymbol \kappa}
\newcommand{\TauB}{\tau_B}

\begin{document}

\title{Dynamics of the effective mass and the anomalous velocity in two-dimensional lattices}
\author{Y. Fang}
\altaffiliation{Current address: Department of Physics, University of California, Berkeley, CA 94720-7300.}
\author{Federico Duque-Gomez}
\email{fduque@physics.utoronto.ca}
\author{J. E. Sipe}
\affiliation{Department of Physics, University of Toronto, Toronto, Ontario, Canada M5S1A7}
\date{\today}

\begin{abstract}
The semiclassical description of the dynamics of wave packets in periodic potentials and subject to an applied force relies on the concepts of effective mass and anomalous transport. This picture is valid if the force changes slowly in time and space, so that the particle described by the wave packet has time to respond according to the properties of the lattice. We analyze the dynamical corrections to this picture when a uniform force is suddenly applied, identifying separate corrections to the usual group and anomalous velocities. We find approximate semianalytical expressions for generalized ``dynamical" group and anomalous velocities, and the associated accelerations. We use a two-dimensional optical lattice with finite Berry curvature to illustrate the semianalytical approximation in a regime where the dynamical corrections are significant, suggesting the possibility of experiments to detect them; we compare the results with a full numerical solution, showing excellent agreement for weak forces.
\end{abstract}


\maketitle

\section{INTRODUCTION}

The bands identified by the Bloch functions of a particle in a periodic potential have both spectral properties, such as their curvature, and topological properties, such as their Berry curvature. Both of these properties appear in the description of the dynamics of a wave packet. The role of the first is probably the more well-known. According to the \emph{effective mass theorem} \cite{Ashcroft76}, a particle moving in the presence of a periodic potential responds to an applied external force $\mbf{F}$ with an \emph{inverse effective mass tensor},  according to 
\begin{equation}
	\frac{d^2}{dt^2} \expval{\posOp{a}} = \expval{\tinveffm{a}{b}{n}{\mbf k}} F^b . \label{E:EffectiveMassTheorem}
\end{equation}
Here $\expval{\posOp{a}}$ is the expectation value of the the $a$th Cartesian component of the wave packet position and repeated Cartesian components, labeled by $b$, are summed over. The expectation value on the right-hand-side is the average over the wave packet of the \emph{local inverse effective mass tensor}, determined by the curvature of the band $n$ with which the wave packet is associated,
\begin{equation}
	\tinveffm{a}{b}{n}{\mbf k} = \frac{1}{\hbar^2} \pdder{k^a}{k^b} \left( \hbar \om{n}{\mbf k} \right), \label{E:KinematicDefEffectiveMass}
\end{equation}
where $\hbar \om{n}{\mbf k}$ is the energy of the band as a function of wave vector $\mathbf{k}$. In the absence of time-reversal or space-inversion symmetry, this simple picture requires an extension due to the link between the dynamics of the particle and the topological properties of the bands \cite{Thouless82, Berry84, Simon83, Hasan10}. Under application of an external force, a wave packet also acquires an additional, \emph{anomalous velocity} \cite{Karplus54, Adams59, Sundaram99, Niu10, Chong10}, proportional to the curl of the applied force and the Berry curvature of the band with which the wave packet is associated, averaged over the wave packet.\\

The connection between the dynamics of wave packets and the spectral and topological properties of bands is thus central to the usual semiclassical description of transport in solid-state physics. However, in statements such as those above, it is important to be precise about what is meant by the ``band with which the wave packet is associated." For a wave packet prepared strictly in one band, the validity of the effective mass theorem and the anomalous transport relies on an adiabatic turning-on of the force \cite{Adams57, Nenciu08}, since the wave packet cannot respond instantaneously according to the properties of the band structure. Upon such an application of the force, the wave packet acquires components of Bloch states of bands other than that which defined the initial wave packet; thus, the ``associated band" has to be taken as that which mainly, but not exclusively, constitutes the wave packet.\\

In contrast to such a scenario, Pfirsch and Spenke showed that if a force is \emph{suddenly} applied, a wave packet strictly in one band responds initially as if the lattice were not present; that is, the ratio of a component of the initial acceleration to the same component of the force is given by the inverse of the bare mass \cite{Pfirsch54}. The response at later times is described by the usual effective mass only on average, as the expectation value of the acceleration oscillates about the usual semiclassical result \cite{Pfirsch54, KriegerIafrate87, HessIafrate88, Iafrate98}. Thus, there are instances when the usual semiclassical expressions fail. Nonetheless, for one-dimensional lattices it has been shown theoretically (see \cite{Duque12} and references therein) that for the sudden application of a force that is not too strong it is possible to define a \emph{dynamical inverse effective mass} of a wave packet, defined as the ratio of the acceleration to the applied force. This quantity is a function of time, initially given by the inverse bare mass when the force is suddenly applied, and oscillating at later times around the usual inverse effective mass, as the wave packet moves through the Brillouin zone. Recently, this dynamical inverse effective mass has been experimentally observed in one-dimensional optical lattices \cite{Rockson14}.\\

This suggests that in lattices of higher dimensionality, where anomalous transport can arise, the breakdown of the effective mass theorem should also be accompanied by corrections to the description of anomalous transport, and that for forces suddenly applied to a wave packet initially restricted to one band it might be possible to identify a \emph{dynamical anomalous velocity} of the wave packet. This would capture the fact that the anomalous velocity should initially vanish when the force is suddenly applied, since the wave packet cannot immediately respond to the lattice, and then eventually oscillate around the usual anomalous velocity that arises in the standard semiclassical description.\\

In this article we extend the description of the dynamics of wave packets subject to suddenly applied forces beyond the one-dimensional case \cite{HessIafrate88, Duque12}, and find that this is indeed so. A semianalytical expression for the expectation value of the velocity is derived using modified Bloch states that decouple the bands in the presence of a uniform force neglecting Zener tunneling \cite{Nenciu91}; these states were introduced by Adams \cite{Adams56, Adams57, Adams59} and studied in more detail by Wannier \cite{Wannier60}. We identify the usual \emph{group velocity}, associated with the inverse effective mass tensor, and the usual anomalous velocity. In addition, we find correction terms to first order in the force, which can be grouped in two terms with different mathematical structure; we interpret one of these terms as an oscillation of the group velocity and the other as an oscillation of the anomalous velocity. Thus we can identify both a \emph{dynamical group velocity} and a \emph{dynamical anomalous velocity} of a wave packet, each of which differs from the usual semiclassical expression by oscillating terms. The time derivative of the sum of these is the expectation value of the acceleration of the wave packet. The time derivative of the dynamical group velocity is related to the applied force by the \emph{dynamical inverse effective mass tensor}, a generalization of the one-dimensional result mentioned above; this tensor is symmetric. The acceleration due to the oscillations in the dynamical anomalous velocity gives rise to an acceleration that is proportional to the force via an antisymmetric tensor; we refer to it as the \emph{dynamical anomalous acceleration} of the wave packet.\\ 

The formalism is illustrated in a two-dimensional example with a tunable honey-comb optical lattice \cite{Tarruell12}; the lowest energy band of this potential has finite local Berry curvature near the Dirac points. We study different trajectories in the Brillouin zone where the oscillations associated with the dynamical quantities mentioned above are significant. We test the validity of our semianalytic approach with a full numerical calculation and show that, for weak forces, our approach is valid over time scales of the order of a Bloch period, as for one-dimensional lattices \cite{Duque12}.\\

Our work is motivated by the recent observation of the dynamics of the effective mass with ultracold atoms in a one-dimensional optical lattice \cite{Rockson14} and the availability of optical lattices of higher dimensionality \cite{Morsch06, Aidelsburger11, Jo12, Tarruell12}. Since optical lattices provide a clean and tunable periodic potential, they are an ideal platform for observing anomalous transport \cite{Dudarev04, Pettini11, Price12}. Furthermore, in this type of lattices the time scale associated with the dynamical oscillations of the effective mass and the anomalous velocity is much longer than in typical solid-state systems, where the period of the oscillations is expected to be of the order of femtoseconds \cite{Pfirsch54, HessIafrate88, Duque12} and the additional scattering due to impurities and phonons make the detection of such oscillations more difficult. However, developments in time-resolved attosecond spectroscopy \cite{Zhu08, Ghimire11, Isanov13, Schubert14} suggest that the observation of electron dynamics in crystals in the sub-femtosecond time scale is possible, opening the possibilities for studying the dynamics of the effective mass and the anomalous velocity presented here.\\

The article is organized as follows. In \refsec{S:TheoreticalFramework} we present the formalism to derive semianalytical expressions for the acceleration and velocity of a wave packet in a lattice of arbitrary dimensionality with an applied force. After introducing our notation in \refsec{S:CMR}, we begin in \refsec{S:EffMassAnomalousTransp} with the modified Bloch states and how wave packets constructed from them, which we call ``Modified Bloch State wave packets," provide the underpinning for the usual semiclassical expressions such as \refeq{E:EffectiveMassTheorem}; in \refsec{S:Dynamics} we show how these expressions are violated when the force is suddenly applied, and how new dynamical quantities associated with the usual semiclassical expressions can be introduced. In \refsec{S:Example} we illustrate the semianalytical formulas with a two-dimensional optical lattice studied earlier by Tarruell et al. \cite{Tarruell12}, and compare the results with a full numerical calculation. Finally, in \refsec{S:Conclusion} we present some conclusions.

%************************

%**RESPONSE OF A CRYSTAL WAVEPACKET TO A UNIFORM FORCE**

\section{RESPONSE OF A CRYSTAL WAVEPACKET TO A UNIFORM FORCE} \label{S:TheoreticalFramework}

%************

%*Crystal momentum representation*

\subsection{Crystal momentum representation} \label{S:CMR}

Consider a particle in a periodic potential, $V(\bpos + \blattvec) = V(\bpos)$, where $\blattvec $ is a lattice vector. In the presence of an external uniform force, $\mbf  F (t)$, the Hamiltonian for the system can be written as
\begin{equation}
	\HOp(t) = \HOp_o - \mbf  F (t) \cdot \bposOp, \label{E:FullHamiltonian}
\end{equation}
where 
\begin{equation}
	\HOp_o \equiv \frac{\bmomOp ^2}{2m} + V(\bposOp) \notag
\end{equation}
is the unperturbed Hamiltonian in terms of the momentum operator, $\bmomOp $, and the bare mass, $m$. The Bloch states that diagonalize $\HOp_o$, 
\begin{equation}
	\HOp_o \ketBloch{n}{\mbf k} = \hbar \om{n}{\mbf k} \ketBloch{n}{\mbf k}, \notag
\end{equation}
have the form
\begin{equation}
	\psi_{n \mbf k}(\bpos) \equiv \rspaceketBloch{n}{\mbf k} = \frac{1}{\sqrt{(2 \pi)^\mathcal{D}}} u_{n \mbf k}(\bpos) e^{i \mbf k \cdot \bpos} \label{E:BlochFunction}
\end{equation}
in real space and are labeled by a band index $n$ and a wave vector $\mbf k$. In \refeq{E:BlochFunction}, $\mathcal{D}$ denotes the dimensionality of the lattice (for example, $\mathcal{D} = 2$ for a two-dimensional lattice) and the function $u_{n \mbf k}(\bpos) \equiv \rspaceket{u_{n \mbf k}}$ has the periodicity of the lattice, $u_{n \mbf k}(\bpos + \blattvec) = u_{n \mbf k}(\bpos)$. In the crystal momentum representation \cite{Blount62}, the state of the particle is written as a wave packet of Bloch states
\begin{equation}
	\ketPsi{t} = \sum_{n} \intBZ c_n(\mbf k, t) \ketBloch{n}{\mbf k}, \notag
\end{equation}
where BZ denotes integration over the (first) Brillouin zone, and the $a$th Cartesian component of the position operator \cite{Blount62}
\begin{multline}
	\meBloch{\posOp{a}}{n_1}{\mbf k_1}{n_2}{\mbf k_2} \equiv \intallpos \psi^{\ast}_{n_1 \mbf k_1}(\bpos) \pos^a \psi_{n_2 \mbf k_2}(\bpos) \\
	= \delta_{n_1 n_2} \left( -i  \pder{k_2^a} \delta(\mbf k_1 - \mbf k_2) \right) + \delta(\mbf k_1 - \mbf k_2) \Lax{a}{n_1}{n_2}{\mbf k_1} \label{E:PosOpCMR}
\end{multline}
is expressed in terms of the matrix elements \cite{Lax74}
\begin{multline}
	\Lax{a}{n_1}{n_2}{\mbf k} \equiv \meUBloch{i \pder{k^a}}{n_1}{\mbf k}{n_2}{\mbf k} \\
	\equiv \intunitcell u^{\ast}_{n_1 \mbf k}(\bpos) i \pder{k^a} u_{n_2 \mbf k}(\bpos).  \label{E:LaxConn}
\end{multline}
Note that the integration in \refeq{E:PosOpCMR} and in any matrix element of the form $\meBloch{\, \cdot \,}{n_1}{\mbf k_1}{n_2}{\mbf k_2}$ is over \emph{all space}; on the other hand, the integration in \refeq{E:LaxConn} and in any matrix element of the form $\meUBloch{\, \cdot \,}{n_1}{\mbf k}{n_2}{\mbf k}$ is over \emph{one unit cell}, with volume $V_{\text{cell}}$.\\

Similarly, for the momentum operator we have
\begin{eqnarray}
	\meBloch{\momOp{a}}{n_1}{\mbf k_1}{n_2}{\mbf k_2} &\equiv& \intallpos \psi^{\ast}_{n_1 \mbf k_1}(\bpos) \frac{\hbar}{i} \pder{\pos^a}  \psi_{n_2 \mbf k_2}(\bpos) \notag \\
	&=& \delta(\mbf k_1 - \mbf k_2) \mommat{a}{n_1}{n_2}{\mbf k_1}, \notag 
\end{eqnarray}
where \cite{Blount62}
\begin{equation}
	\mommat{a}{n_1}{n_2}{\mbf k} \equiv \delta_{n_1 n_2} \hbar k^a + \meUBloch{\momOp{a}}{n_1}{\mbf k}{n_2}{\mbf k}. \notag
\end{equation} 
For non-degenerate bands, the off-diagonal elements of \refeq{E:LaxConn} are related to the momentum matrix elements by \cite{Lax74}
\begin{equation}
	\Lax{a}{n_1}{n_2}{\mbf k} = \frac{1}{i m} \frac{\mommat{a}{n_1}{n_2}{\mbf k}}{\om{n_1 n_2}{\mbf k}} \text{ (for $\om{n_1}{\mbf k} \ne \om{n_2}{\mbf k}$)}, \notag
\end{equation}
where $\om{n_1 n_2}{\mbf k} \equiv \om{n_1}{\mbf k} - \om{n_2}{\mbf k}$. The diagonal element $\bLax{n}{n}{\mbf k}$ is known as the \emph{Berry connection}. Despite being ``gauge-dependent,'' in that it depends on how the phases of the Bloch states are set throughout the Brillouin zone, the Berry connection has dynamical consequences through the \emph{Berry curvature} \cite{Karplus54, Adams59, Sundaram99, Niu10}, which is gauge-independent. The local Berry curvature, $\bbryc{n}{\mbf k}$, has components given by
\begin{equation}
	\bryc{l}{n}{\mbf k} \equiv \levic{l}{a}{b} \pder{k^a} \Lax{b}{n}{n}{\mbf k}, \label{E:DefBerryCurvature} 
\end{equation}
where $\levic{l}{a}{b}$ is the antisymmetric Levi-Civita symbol, and repeated Cartesian components are summed over.

%************

%*The effective mass theorem and  anomalous transport*
\subsection{Modified Bloch states, the effective mass theorem, and anomalous transport} \label{S:EffMassAnomalousTransp}

For a wave packet subjected to a force $\mbf F(t)$ (see \refeq{E:FullHamiltonian}) and described by a ket $\ketPsi{t}$, the expectation value of the acceleration follows from Ehrenfest's theorem,
\begin{eqnarray}
	\expval{\baccOp(t)} &\equiv& \frac{d^2}{d t^2} \expvalS{\bposOp}{t} \notag \\
	&=& \frac{\mbf  F(t)}{m} + \frac{1}{i \hbar m} \expvalS{\comm{\bmomOp}{\HOp_o}}{t}. \label{E:EhrenfestAccel}
\end{eqnarray}
Suppose now we consider a force that turns on at $t=0$, $\mbf F (t) = \mbf 0$ for $t<0$ and $\mbf F (t) = \mbf F$ for $t \ge 0$. If initially the wave packet is formed only by Bloch states from a single band, say band $N$,
\begin{equation}
	\ketPsi{0} = \ket{\bar{\psi}_N} \equiv \intBZ f_N(\mbf k) \ketBloch{N}{\mbf k}, \label{E:InitialWvp}
\end{equation}
at $t=0$ the commutator in \refeq{E:EhrenfestAccel} vanishes and we have $\expval{\baccOp(0^+)} = \mbf F/m$. That is, the wave packet responds initially with the inverse \emph{bare} mass of the particle and not the inverse effective mass tensor. This is an old result \cite{Pfirsch54, KriegerIafrate87, Iafrate98, HessIafrate88}: a wave packet formed by Bloch states from a single band does not respond to the periodic potential of the lattice when the force is applied, and at very early times the particle generally responds to an applied force as if it were free. At later times the wave packet cannot remain in one band only, as the force inevitably couples the Bloch states of different bands, and it is through this coupling that the lattice makes itself felt. However, we can still define a wave packet that remains \emph{mainly} in one band and responds essentially with the properties of that band; the amplitudes of such a wave packet in neighboring bands are due to the interband mixing induced by the force.\\

In a treatment that was later shown to be correct with the neglect of Zener tunneling \cite{Nenciu91}, Wannier found that for a constant and uniform force, $\mbf F(t) = \mbf F$, the interband mixing can be captured by \emph{modified Bloch states}, $\ketModBloch{n}{\mbf k}$, that are related to the original Bloch states $\ketBloch{n}{\mbf k}$ by a unitary transformation \cite{Wannier60},
\begin{equation}
	\ketModBloch{n}{\mbf k} \equiv \sum_{n'} \ketBloch{n'}{\mbf k} U_{n' n}(\mbf k). \label{E:DefModBlochStates}
\end{equation}
The unitary transformation $U_{n' n}(\mbf k)$ can be constructed to different orders in the force \cite{Wannier60, Duque12}; in the first order approximation and assuming no degeneracies \footnote{In \cite{Wannier60}, Wannier also considers the generalization of his decoupling method to bands that share degenerate points. For simplicity, we assume nondegenerate bands throughout, an assumption that is valid in the type of potential considered in \refsec{S:Example}}, these modified Bloch states take the form
\begin{equation}
	\ketModBloch{n}{\mbf k} \approx \ketBloch{n}{\mbf k} + \sum_{n'} \ketBloch{n'}{\mbf k} \foMix_{n' n}(\mbf k), \notag
\end{equation}
where $\foMix(\mbf k)$ is an off-diagonal matrix with elements \cite{Adams56, Duque12}
\begin{equation}
	\foMix_{n_1 n_2}(\mbf k) \equiv \frac{F^b \Lax{b}{n_1}{n_2}{\mbf k}}{\hbar \om{n_1 n_2}{\mbf k}} (1-\delta_{n_1 n_2}). \notag
\end{equation}
As a result of this construction, a wave packet formed only by modified Bloch states associated with band $N$,
\begin{equation}
	\ketWvpMB{N}{0} = \intBZ \bar{b}_N(\mbf k) \ketModBloch{N}{\mbf k}, \notag
\end{equation}
evolves in time moving through the Brillouin zone without mixing with other modified Bloch states from neighboring bands. To first order in the force, we find \cite{Wannier60, Duque12}
\begin{equation}
	\ketWvpMB{N}{t} = \intBZ \bar{b}_N(\bkmov) e^{-i \gphase{N}{\bkmov}{t}} \ketModBloch{N}{\mbf k}, \label{E:OneModBandWvp}
\end{equation}
where $\bkmov \equiv \mbf k - \mbf F t/\hbar$ and
\begin{equation}
	\gphase{n}{\mbf k}{t} \equiv \int_{0}^{t} \left[ \om{n}{\mbf k + \frac{1}{\hbar} \mathbf{F} t'} - \frac{1}{\hbar} \mathbf{F} \cdot \bLax{n}{n}{\mbf k + \frac{1}{\hbar} \mathbf{F} t'} \right] dt'. \notag
\end{equation}
We refer to the wave packet $\ketWvpMB{N}{t}$ as a MBS wave packet, where MBS stands for Modified Bloch State. This type of wave packet satisfies the effective mass theorem \emph{at all times}. This can be shown by using the matrix elements \cite{Duque12} 
\begin{equation}
	\modforce{a}{n_1}{n_2}{\mbf k} \equiv i \sum_{n_1', n_2'} U^{\dagger}_{n_1 n_1'}(\mbf k)  \mommat{a}{n_1'}{n_2'}{\mbf k} \om{n_1' n_2'}{\mbf k} U_{n_2' n_2}(\mbf k) \notag
\end{equation}
to rewrite \refeq{E:EhrenfestAccel} for $\ketWvpMB{N}{t}$ as
\begin{equation}
	\frac{d^2}{d t^2} \expvalWvpMB{\posOp{a}}{N}{t} = \frac{F^a}{m} + \intBZ |\bar{b}_N(\bkmov)|^2 \modforce{a}{N}{N}{\mbf k}. \label{E:AccelerationModBlochWvp}
\end{equation}
To first order in the force, we can write
\begin{equation}
	\modforce{a}{N}{N}{\mbf k} \approx \frac{1}{m} \sum_{n \ne N} \frac{\mommat{a}{N}{n}{\mbf k} \mommat{b}{n}{N}{\mbf k} + \mommat{b}{N}{n}{\mbf k} \mommat{a}{n}{N}{\mbf k}}{\hbar \om{N n}{\mbf k}} F^b, \label{E:LatticeForce}
\end{equation}
and the sum rule for the local inverse effective mass tensor \cite{Lax74},
\begin{multline}
	m \tinveffm{a}{b}{N}{\mbf k} - \delta^{ab} = \\
	\frac{1}{m} \sum_{n \ne N} \frac{\mommat{a}{N}{n}{\mbf k} \mommat{b}{n}{N}{\mbf k} + \mommat{b}{N}{n}{\mbf k} \mommat{a}{n}{N}{\mbf k}}{\hbar \om{N n}{\mbf k}}, \label{E:EffectiveMassSumRule}
\end{multline}
reduces \refeq{E:AccelerationModBlochWvp} to
\begin{multline}
	\frac{d^2}{d t^2} \expvalWvpMB{\posOp{a}}{N}{t} = \\
	\intBZ |\bar{b}_N(\bkmov)|^2 \tinveffm{a}{b}{N}{\mbf k} F^b, \label{E:EffMassThmModBlochWvp}
\end{multline}
in accordance with the effective mass theorem, \refeq{E:EffectiveMassTheorem}. Note that, compared with \refeq{E:KinematicDefEffectiveMass}, the sum rule in \refeq{E:EffectiveMassSumRule} reveals the truly multi-band nature of the local inverse effective mass tensor \footnote{\refeq{E:KinematicDefEffectiveMass} or, equivalently \refeq{E:EffectiveMassSumRule}, are also used as definitions of the effective mass in other scenarios in solid-state physics, such as in the $\mbf k \cdot \mbf p$ method for calculating band structures \cite{Lax74}.}, which is reflected in the use of modified Bloch states to construct a wave packet that responds with the effective mass \emph{at all times}.\\ 

We can follow the same strategy for the expectation value of the velocity. We find
\begin{equation}
	\expval{\bvelOp(t)} \equiv \frac{d}{d t} \expvalS{\bposOp}{t} = \frac{1}{m} \expvalS{\bmomOp}{t} \label{E:EhrenfestVel}
\end{equation}
(see \refeq{E:EhrenfestAccel}). In the particular case of a MBS wave packet, $\ketPsi{t} = \ketWvpMB{N}{t}$, \refeq{E:EhrenfestVel} becomes
\begin{equation}
	\frac{d}{d t} \expvalWvpMB{\posOp{a}}{N}{t} = \intBZ |\bar{b}_N(\bkmov)|^2 \frac{1}{m} \modmom{a}{N}{N}{\mbf k}, \label{E:VelocityModBlochWvp}
\end{equation}
where $\modmom{a}{n_1}{n_2}{\mbf k}$ are the transformed momentum matrix elements,
\begin{equation}
	\modmom{a}{n_1}{n_2}{\mbf k} \equiv \sum_{n_1', n_2'} U^{\dagger}_{n_1 n_1'}(\mbf k)  \mommat{a}{n_1'}{n_2'}{\mbf k} U_{n_2' n_2}(\mbf k). \notag
\end{equation}
Analogously to \refeq{E:LatticeForce}, the diagonal elements $\modmom{a}{N}{N}{\mbf k}$ have a simple form to first order in the force,
\begin{equation}
	\frac{1}{m}\modmom{a}{N}{N}{\mbf k} \approx \vg{a}{N}{\mbf k} + \frac{1}{\hbar} \tv{a}{b}{N}{N}{\mbf k} F^b. \label{E:DiagTransfMom}
\end{equation}
The first term in this expression corresponds to the \emph{local group velocity} \cite{Lax74}, 
\begin{equation}
	\vg{a}{N}{\mbf k} \equiv \frac{\mommat{a}{N}{N}{\mbf k}}{m} = \frac{1}{\hbar} \pder{k^a} (\hbar \om{N}{\mbf k}), \label{E:GroupVel}
\end{equation}
given by the gradient of the band energy; this term is directly related to the local inverse effective mass tensor since
\begin{multline}
	\frac{d}{dt}  \intBZ |\bar{b}_N(\bkmov)|^2 \, \vg{a}{N}{\mbf k} = \\
	- \intBZ \left( \pder{k^b} |\bar{b}_N(\bkmov)|^2 \right) \frac{1}{\hbar^2} \pder{k^a} (\hbar \om{N}{\mbf k}) F^b =\\
	\intBZ |\bar{b}_N(\bkmov)|^2 \, \tinveffm{a}{b}{n}{\mbf k} F^b. \label{E:AccelerationGroupVel}
\end{multline}
The contribution from the first term on the right-hand-side of \refeq{E:DiagTransfMom} to \refeq{E:VelocityModBlochWvp} is not surprising; as a wave packet simply moves through the Brillouin zone at a pace proportional to the force, it acquires the group velocity (see \refeq{E:GroupVel}) associated with the band energy. The second term on the right-hand-side of \refeq{E:DiagTransfMom}, on the other hand, has a completely different structure given by the antisymmetric tensor 
\begin{equation}
	\tv{a}{b}{N}{N}{\mbf k} \equiv \sum_{n \ne N} 2 \, \text{Im} \left[ \Lax{a}{N}{n}{\mbf k} \Lax{b}{n}{N}{\mbf k} \right]. \label{E:AntisymTensAnVel}
\end{equation}
As a result of the sum rule \cite{Lax74}
\begin{equation}
	\levic{a}{l}{b} \bryc{l}{N}{\mbf k} = \sum_{n \ne N} 2 \, \text{Im} \left[ \Lax{a}{N}{n}{\mbf k} \Lax{b}{n}{N}{\mbf k} \right], \notag
\end{equation}
the second term on the right-hand-side of \refeq{E:DiagTransfMom} becomes the \emph{local anomalous velocity} \cite{Adams59},
\begin{equation}
	\van{a}{N}{\mbf k} \equiv \frac{1}{\hbar} \levic{a}{l}{b} \bryc{l}{N}{\mbf k} F^b, \label{E:DefAnomalousVel}
\end{equation}
a first order correction to the velocity associated with the local Berry curvature, \refeq{E:DefBerryCurvature}. Thus, we can rewrite \refeq{E:VelocityModBlochWvp} as
\begin{multline}
	\frac{d}{d t} \expvalWvpMB{\posOp{a}}{N}{t} =\\
	\intBZ |\bar{b}_N(\bkmov)|^2 \left( \vg{a}{N}{\mbf k} + \van{a}{N}{\mbf k} \right),\label{E:UsualVelModBlochWvp}
\end{multline}
including both the group velocity and the anomalous velocity of the wave packet. These velocities are periodic in time as the wave packet traverses the Brillouin zone and returns to its starting point in reciprocal space; the period of this motion in reciprocal space is the Bloch period, $\tau_B$ \cite{Bloch29}. The anomalous transport correction to the usual effective mass behavior is only important for potentials that break space-inversion symmetry or time-reversal symmetry, where the local Berry curvature is different from zero \footnote{This is strictly true if there are no degeneracies. For example, the two-dimensional hexagonal lattice with six-fold symmetry, characteristic of systems such as graphene, has singular local Berry curvature at the Dirac points where the two lowest bands touch (see \cite{Fuchs10}).}. Note that if we calculate the time derivative of \refeq{E:UsualVelModBlochWvp} and keep terms linear in the force, we get the same result as \refeq{E:EffMassThmModBlochWvp} since the anomalous velocity term does not contribute to first order in the force, while the group velocity term contributes the acceleration described by the usual inverse effective mass tensor (see \refeq{E:AccelerationGroupVel}).

%************

%*Dynamics of the effective mass and the anomalous velocity*
\subsection{Dynamics of the effective mass and the anomalous velocity} \label{S:Dynamics}

We can employ the modified Bloch states to describe the motion of a wave packet, originally consisting of a superposition of usual Bloch states from band $N$, \refeq{E:InitialWvp}, when a force is applied. Were the force increased \emph{adiabatically} from zero to $\mbf F$, the wave packet would acquire the particular composition required to move according to \refeqs{E:EffMassThmModBlochWvp} and \eqref{E:UsualVelModBlochWvp} \cite{Adams57, Nenciu08}. But if the force is applied instantaneously, for $t>0$ we can seek an expansion of the ket $\ketPsi{t}$ as a superposition of MBS wave packets (see \refeq{E:OneModBandWvp}), with coefficients chosen so that at $t=0$ the state yields \refeq{E:InitialWvp}. To linear order in the force, which is the order to which we have constructed the $\ketModBloch{n}{\mbf k}$, this can be easily done; to zeroth order $\ketPsi{t}$ will just be the appropriate $\ketWvpMB{N}{t}$, and to first order there will be contributions from other $\ketWvpMB{n}{t}$ with $n \ne N$ \cite{Duque12}. In this scenario we will see that there is a transition from the bare mass response to the semiclassical dynamics described with the local inverse effective mass tensor, \refeq{E:KinematicDefEffectiveMass}, and the local anomalous velocity, \refeq{E:DefAnomalousVel}; the transition is characterized by oscillations of the acceleration and the velocity around the values predicted by \refeqs{E:EffMassThmModBlochWvp} and \eqref{E:UsualVelModBlochWvp}. We now establish those dynamics.\\

Following \cite{Duque12}, the ket for $t \ge 0$ is
\begin{equation}
	\ketPsi{t} = \sum_{n} \intBZ b_n(\mbf k, t) \ketModBloch{n}{\mbf k}, \label{E:OurWavepacket}
\end{equation}
where to first order in the force
\begin{eqnarray}
	b_N(\mbf k, t) &=& f_N(\bkmov) e^{-i \gphase{N}{\bkmov}{t}}, \label{E:AmplitudeMain} \\
	b_n(\mbf k, t) &=& -f_N(\bkmov) \foMix_{n N}(\bkmov) e^{-i \gphase{n}{\bkmov}{t}}. \label{E:AmplitudeNeighbor}
\end{eqnarray}
The expectation value of the velocity,
\begin{equation} 
	\expval{\velOp^a(t)} = \frac{1}{m} \sum_{n_1, n_2} \intBZ b^{\ast}_{n_1}(\mbf k, t) b_{n_2}(\mbf k, t) \modmom{a}{n_1}{n_2}{\mbf k}, \notag
\end{equation}
can be split in three terms,
\begin{multline}
	\expval{\velOp ^a(t)} \approx \\
	\intBZ |f_{N}(\bkmov)|^2 \left( \vg{a}{N}{\mbf k} + \van{a}{N}{\mbf k} + \vosc{a}{N}{\mbf k}{t} \right). \label{E:VelocityThreeTerms}
\end{multline}
As expected, the first two contributions involve the usual group velocity from \refeq{E:GroupVel}, and the anomalous velocity from \refeq{E:DefAnomalousVel}. In the additional term, $\vosc{a}{N}{\mbf k}{t}$ can be written as
\begin{equation}
	\vosc{a}{N}{\mbf k}{t} = -\frac{1}{\hbar} \tosc{a}{b}{N}{\mbf k}{t} F^b, \notag
\end{equation}
where the tensor
\begin{equation}
	\tosc{a}{b}{N}{\mbf k}{t} \equiv \sum_{n \ne N} \frac{\om{nN}{\mbf k}}{\om{nN}{\bkmov}} 2 \, \text{Im} \left[ \Lax{a}{N}{n}{\mbf k} \Lax{b}{n}{N}{\bkmov} e^{-i \gamma_{n N}(\bkmov, t)} \right] \label{E:OscillationTensor}
\end{equation}
is a time-dependent generalization of \refeq{E:AntisymTensAnVel} with $\gamma_{n_1 n_2}(\mbf k, t) \equiv \gamma_{n_1}(\mbf k, t) - \gamma_{n_2}(\mbf k, t)$. We will see that $\vosc{a}{N}{\mbf k}{t}$ describes oscillations of $\expval{\velOp ^a(t)}$ about the result \refeq{E:UsualVelModBlochWvp}. At $t=0$, $\tosc{a}{b}{N}{\mbf k}{t}$ reduces to the antisymmetric tensor $\tv{a}{b}{N}{N}{\mbf k}$, but for $t>0$ it does not have definite symmetry. Nevertheless, we can still decompose $\tosc{a}{b}{N}{\mbf k}{t}$ uniquely into symmetric and antisymmetric parts,
\begin{equation}
	\tosc{a}{b}{N}{\mbf k}{t} = \stosc{a}{b}{N}{\mbf k}{t} + \atosc{a}{b}{N}{\mbf k}{t},
\end{equation}
where
\begin{eqnarray}
	\stosc{a}{b}{N}{\mbf k}{t} &\equiv& \frac{1}{2} \left( \tosc{a}{b}{N}{\mbf k}{t} + \tosc{b}{a}{N}{\mbf k}{t} \right), \label{E:SymmTensVosc} \\
	\atosc{a}{b}{N}{\mbf k}{t} &\equiv& \frac{1}{2} \left( \tosc{a}{b}{N}{\mbf k}{t} - \tosc{b}{a}{N}{\mbf k}{t} \right). \label{E:AntisymmTensVosc}
\end{eqnarray}
The antisymmetric part, \refeq{E:AntisymmTensVosc}, can also be expressed in terms of the axial-vector $\bbrycosc{N}{\mbf k}{t}$ with components
\begin{equation}
	\brycosc{l}{N}{\mbf k}{t} \equiv - \frac{1}{2} \levic{l}{a}{b} \atosc{a}{b}{N}{\mbf k}{t}, \label{E:BerryCurvOsc}
\end{equation}
so that
\begin{equation}
	\atosc{a}{b}{N}{\mbf k}{t} = \levic{a}{l}{b} \brycosc{l}{N}{\mbf k}{t}. \notag
\end{equation}
Defining
\begin{multline}
	\LaxSq{\pm}{a}{b}{N}{n}{\mbf k}{t} \equiv \Lax{a}{N}{n}{\mbf k} \Lax{b}{n}{N}{\bkmov} e^{-i \gamma_{nN}(\bkmov, t)} \\
	\pm \Lax{a}{N}{n}{\bkmov} e^{-i \gamma_{Nn}(\bkmov, t)} \Lax{b}{n}{N}{\mbf k}, \notag
\end{multline}
\refeqs{E:SymmTensVosc} and \eqref{E:BerryCurvOsc} become
\begin{equation}
	\stosc{a}{b}{N}{\mbf k}{t} =  \sum_{n \ne N} \text{Im} \left[ \LaxSq{-}{a}{b}{N}{n}{\mbf k}{t} \right] \frac{\om{n N}{\mbf k}}{\om{n N}{\bkmov}} \label{E:ExplicitSymmTensVosc}
\end{equation}
and
\begin{equation}
	\brycosc{l}{N}{\mbf k}{t} = - \frac{1}{2} \levic{l}{a}{b} \sum_{n \ne N} \text{Im} \left[ \LaxSq{+}{a}{b}{N}{n}{\mbf k}{t} \right] \frac{\om{n N}{\mbf k}}{\om{n N}{\bkmov}}, \label{E:ExplicitBerryCurvOsc}
\end{equation}
respectively.\\

%Figure 1
\begin{figure}
 \centering
 \includegraphics[width=\columnwidth]{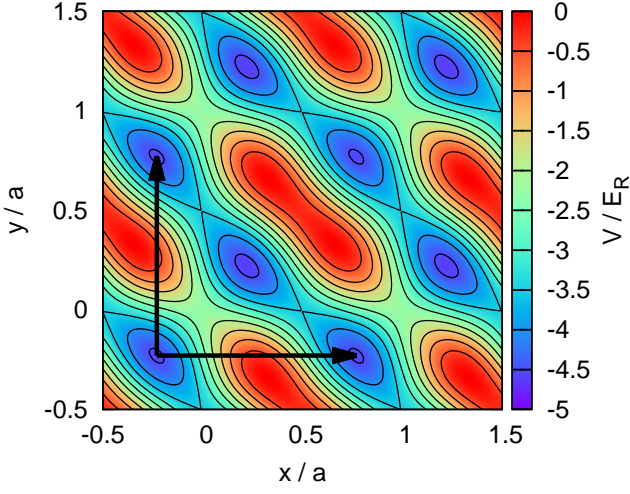}
 \caption{(Color online) Honeycomb lattice given by \refeq{E:Latticepotential}. The x- and y-axes give the position, in units of $a$. The color plot shows the value of the potential in units of the recoil energy $E_R$. Two lattice vectors are also shown with black arrows.}
 \label{fig:Potential}
\end{figure}

With this formal decomposition in hand, we link some of the dynamical oscillations described by $\vosc{a}{N}{\mbf k}{t}$ with the group velocity and some with the anomalous velocity by writing the expectation value \refeq{E:VelocityThreeTerms} of the velocity $\expval{\velOp ^a(t)}$ as 
\begin{equation}
	\expval{\velOp ^a(t)} \approx V^{\text{g},a}(t) + V^{\text{an},a}(t), \label{E:vuse}
\end{equation}
where we identify the \emph{dynamical group velocity} $\mbf V^{\text{g}}(t)$ of the wave packet, with components 
\begin{equation}
	V^{\text{g},a}(t) \equiv \intBZ |f_N(\bkmov)|^2 \left( \vg{a}{N}{\mbf k} - \frac{1}{\hbar} \stosc{a}{b}{N}{\mbf k}{t} F^b \right), \label{E:Vga} 
\end{equation}
and the \emph{dynamical anomalous velocity} $\mbf V^{\text{an}}(t)$ of the wave packet, with components
\begin{equation}
	V^{\text{an},a}(t) \equiv \intBZ |f_N(\bkmov)|^2 \, \left( \van{a}{N}{\mbf k} -\frac{1}{\hbar}\atosc{a}{b}{N}{\mbf k}{t} F^b \right). \label{E:Vana}
\end{equation}
The leading term in parentheses in the integrands from \refeqs{E:Vga} and \eqref{E:Vana} are the contributions that would be expected from a MBS wave packet associated with band $N$ (see \refeq{E:UsualVelModBlochWvp}); the other terms in parentheses give oscillatory corrections. Note that \refeq{E:Vana} can be written as
\begin{equation}
	V^{\text{an},a}(t) = \frac{1}{\hbar} \levic{a}{l}{b} \brycwvp{l}{N}{t} F^b, \label{E:DynAnomVel}
\end{equation}
where
\begin{equation}
	\bbrycwvp{N}{t} \equiv \intBZ |f_{N}(\bkmov)|^2 \left( \bbryc{N}{\mbf k} - \bbrycosc{N}{\mbf k}{t} \right) \notag
\end{equation}
can be interpreted as the \emph{dynamical Berry curvature} ``seen" by the wave packet as it moves through the Brillouin zone.\\

%Figure 2
\begin{figure}
 \centering
 \includegraphics[width=\columnwidth]{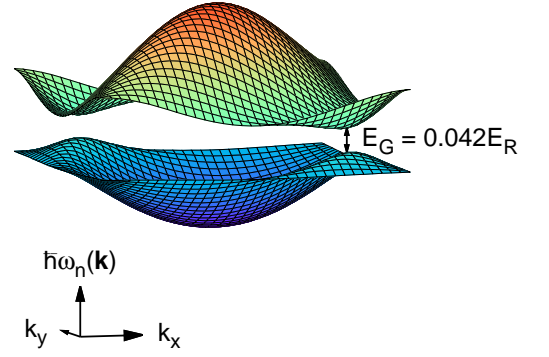}
 \caption{(Color online) Energy spectrum of the first two bands. When $\theta = 1.02\pi$, a band gap of size $E_G = 0.042E_R$ is opened at the two Dirac points, where the two bands meet when $\theta = \pi$.}
  \label{fig:Band}
\end{figure}

The identification of the term involving $\stosc{a}{b}{N}{\mbf k}{t}$ with the group velocity of the wave packet is justified by the fact that the time derivative of $\mbf V^{\text{g}}(t)$,
\begin{equation}
	\mbf A^{\text{g}}(t) \equiv \frac{d}{dt} \mbf V^{\text{g}}(t), \notag
\end{equation}
is found to have components
\begin{equation}
	A^{\text{g}, a}(t) = \left[ \frac{1}{M^{\ast}(t)} \right]^{ab} F^b, \notag
\end{equation}
where we have introduced a \emph{dynamical inverse effective mass tensor},
\begin{multline}
	\left[ \frac{1}{M^{\ast}(t)} \right]^{ab} \equiv \\
	\intBZ |f_{N}(\bkmov)|^2 \,  \left( \tinveffm{a}{b}{N}{\mbf k} - \frac{1}{m} \staosc{a}{b}{N}{\mbf k}{t} \right), \label{E:Mstar}
\end{multline}
and where the tensor
\begin{equation}
	\staosc{a}{b}{N}{\mbf k}{t} \equiv - \frac{m}{\hbar} \sum_{n \ne N} \text{Re} \left[ \LaxSq{+}{a}{b}{N}{n}{\mbf k}{t} \right] \frac{(\om{n N}{\mbf k})^2}{\om{n N}{\bkmov}}, \label{E:SymmAccelTensor}
\end{equation}
is \emph{symmetric} with respect to its Cartesian components in the same way as the local inverse effective mass tensor \refeq{E:KinematicDefEffectiveMass}, leading to a dynamical inverse effective mass tensor that is symmetric. The first contribution in parentheses in the integrand of \refeq{E:Mstar} gives the result that would be expected for a MBS wave packet (see \refeq{E:EffMassThmModBlochWvp}); the term involving $\staosc{a}{b}{N}{\mbf k}{t}$ describes oscillatory corrections. Within the approximation of \refeq{E:vuse}, the \emph{full} acceleration of the wave packet,
\begin{equation}
	\expval{\baccOp(t)} = \frac{d}{dt} \expval{\bvelOp(t)} = \mbf A^{\text{g}}(t) + \mbf A^{\text{an}}(t) \label{E:DecompAcceleration}
\end{equation}
includes a \emph{dynamical anomalous acceleration} $\mbf A^{\text{an}}(t)$, which is the time derivative of $\mbf V^{\text{an}}(t)$. This dynamical anomalous acceleration is perpendicular to the force, and it has components
\begin{equation}
	A^{\text{an},a}(t) = \intBZ |f_{N}(\bkmov)|^2 \, \left(- \frac{1}{m} \ataosc{a}{b}{N}{\mbf k}{t} \right) F^b, \notag
\end{equation}
given by the \emph{antisymmetric} tensor
\begin{equation}
	\ataosc{a}{b}{N}{\mbf k}{t} \equiv \levic{a}{l}{b} \axvaosc{l}{N}{\mbf k}{t}, \label{E:AntisymmAccelTensor}
\end{equation} 
where
\begin{equation}
	\axvaosc{l}{N}{\mbf k}{t} \equiv \frac{m}{2\hbar} \levic{l}{a}{b} \sum_{n \ne N} \text{Re} \left[ \LaxSq{-}{a}{b}{N}{n}{\mbf k}{t} \right] \frac{(\om{n N}{\mbf k})^2}{\om{n N}{\bkmov}} \notag
\end{equation}
are the components of an axial-vector. Unlike the dynamical inverse effective mass tensor and the acceleration it describes, to first order in the force the dynamical anomalous acceleration contains only oscillatory terms (see the discussion after \refeq{E:UsualVelModBlochWvp}). Note that the acceleration in \refeq{E:DecompAcceleration} can also be derived directly from Ehrenfest's theorem using \refeq{E:EhrenfestAccel} with the wave packet \refeq{E:OurWavepacket}.\\

%Figure 3
\begin{figure}
 \centering
 \includegraphics[width=\columnwidth]{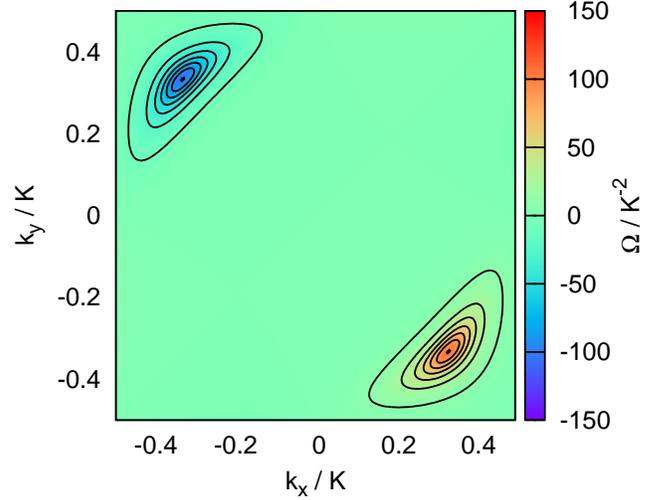}
 \caption{(Color online) Local Berry curvature of the lowest band ($n=1$). The x- and y-axes give the location in the Brillouin zone while the color denotes the value of the local Berry curvature.}
 \label{fig:Berry}
\end{figure}

At the initial time, \refeqs{E:ExplicitSymmTensVosc} and \eqref{E:ExplicitBerryCurvOsc} reduce to
\begin{equation}
	\stosc{a}{b}{N}{\mbf k}{0} = 0 \, \text{ and } \,  \brycosc{l}{N}{\mbf k}{0} = \bryc{l}{N}{\mbf k}, \notag
\end{equation}
and so
\begin{equation}
	V^{\text{g},a}(0) = \intBZ |f_N(\mbf k)|^2 \, \vg{a}{N}{\mbf k} \, \text{ and } \, V^{\text{an},a}(0) = 0, \label{E:InitialGroupAnomVel}
\end{equation}
consistently with the initial behavior found using Ehrenfest's theorem for the initial state \refeq{E:InitialWvp}; initially the Berry curvature ``seen" by the wave packet vanishes, and there is no anomalous velocity. For $t>0$ the wave packet's velocity acquires dynamical oscillations around the usual group velocity, in a way similar to the earlier results found for one-dimensional lattices \cite{Duque12}. Since at the initial time
\begin{equation}
	\staosc{a}{b}{N}{\mbf k}{0} = m \tinveffm{a}{b}{N}{\mbf k} - \delta^{ab} \, \text{ and } \, \ataosc{a}{b}{N}{\mbf k}{0} = 0 \notag
\end{equation}
(see \refeqs{E:EffectiveMassSumRule}, \eqref{E:SymmAccelTensor} and \eqref{E:AntisymmAccelTensor}), the dynamical inverse effective mass tensor is initially given by the inverse bare mass,
\begin{equation}
	\left[ \frac{1}{M^{\ast}(0)} \right]^{ab} = \frac{\delta^{ab}}{m} \, \text{ and } \, A^{\text{an}, a}(0) = 0. \label{E:InitialInvEffMassTensor}
\end{equation}
Hence, the particle initially responds with the bare mass \cite{Pfirsch54, KriegerIafrate87, Duque12}; afterwards, the dynamical inverse effective mass tensor acquires oscillations about the usual inverse effective mass tensor. In addition, the anomalous transport described by $\mbf V^{\text{an}}(t)$ has its own dynamics; rather than just the Berry curvature, it is governed by the dynamical Berry curvature $\brycwvp{l}{N}{t}$, which contains its own oscillatory terms.\\

%Figure 4
\begin{figure}
 \centering
 \includegraphics[width=\columnwidth]{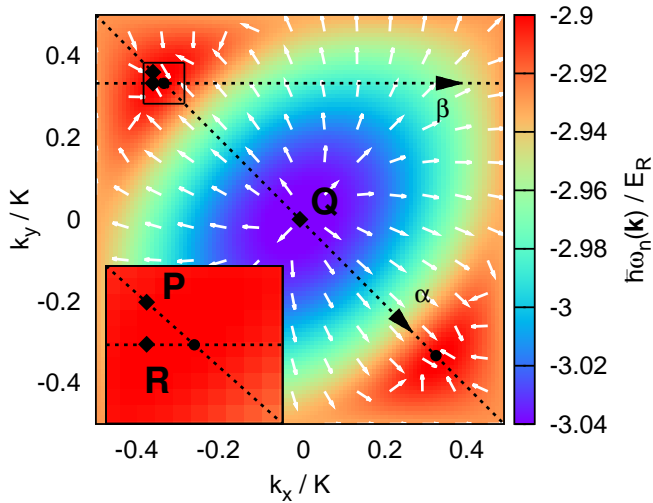}
 \caption{(Color online) Two paths $\alpha$ and $\beta$ in the first energy band ($n=1$). The direction of the energy gradient, proportional to the local group velocity, is shown with the vector field. Both paths pass through at least one of the Dirac points, shown with black dots. For the diagonal path $\alpha$, which is highly symmetric, the local group velocity is always aligned with the path; for the horizontal path $\beta$ this is not so. Three trajectories with different starting points (indicated by black diamonds) are discussed here. In the first one, the wave packet is prepared at point \textbf{P} and travels along $\alpha$ in the direction indicated by the arrow. In the second one, the wave packet is prepared at \textbf{Q} and takes the same path. In the last trajectory, the wave packet starts at \textbf{R} and travels along $\beta$. The inset magnifies the region near \textbf{P} and \textbf{R} to clarify the difference between the two.}
 \label{fig:trajectories}
\end{figure}

%************************

%**EXAMPLE**

\section{EXAMPLE} \label{S:Example}

We now apply the semianalytical expressions for the dynamics of wave packets to a two-dimensional optical lattice. We begin with the band structure and local Berry curvature of the lattice we consider (\refsec{S:2DLattice}). A wave packet built in the first band with a Gaussian envelope function is then set along different trajectories in the Brillouin zone; its dynamics are calculated through the semianalytical approach (\refsec{S:SemianalyticalResults}). Finally, this result is compared with a full numerical solution (\refsec{S:Numerical}).\\

%Figure 5
\begin{figure}
\centering
\includegraphics[width=\columnwidth]{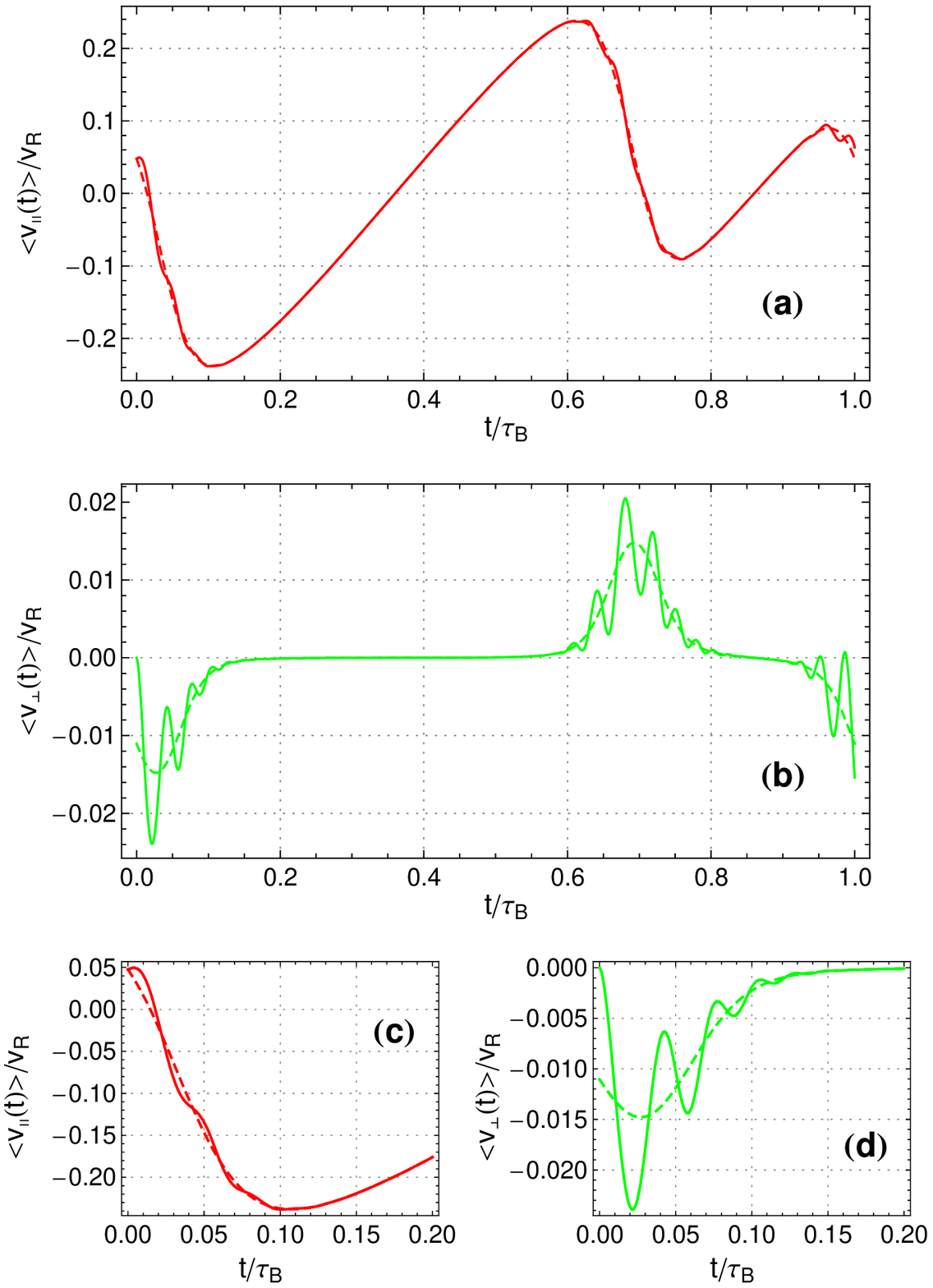}
\caption{(Color online) Expectation value of the velocity from the semianalytical approximation, \refeq{E:VelocityDecomp}, without the oscillating term $\bvoscint{t}$ (dashed lines) and including it (solid lines). In this example the force points in the $(\hat{\mbf x} - \hat{\mbf y})/\sqrt{2}$ direction with $\tilde{F}=1/2000$; the wave packet moves through the Brillouin zone starting at point \textbf{P} and following path $\alpha$ (see \reffig{fig:trajectories}). The red curves correspond to $\expval{\text{v}_{\parallel}(t)}$, the component of the velocity parallel to the force, while the green curves correspond to $\expval{\text{v}_{\perp}(t)}$, the component of the velocity perpendicular to the force ($(\hat{\mbf x} + \hat{\mbf y})/\sqrt{2}$ direction). Note that $\bveffint{t}$ and $\bvanint{t}$ only contribute to the parallel and perpendicular directions, respectively. (a-b) Results over one Bloch period, $\tau_B = \hbar K / F$. (c-d) Initial behavior of the velocity, showing the oscillations due to $\bvoscint{t}$.}
\label{fig:DiagPath}
\end{figure}

%************

%*Two-dimensional optical lattice*

\subsection{Two-dimensional optical lattice}\label{S:2DLattice}

We consider the optical lattice described by Tarruell et al. \cite{Tarruell12}, created by interference of three retro-reflected laser beams, which leads to a potential 
\begin{multline} 
	V(x,y) = \frac{-V_{\bar{X}}}{2} \cos\big(K \,(x+y) + \theta\big) - \frac{V_X}{2} \cos\big(K \, (x+y)\big) \\
	- \frac{V_Y}{2}\cos\big(K \, (x-y)\big) - \sqrt{V_X V_Y}\big( \cos(Kx) + \cos(Ky) \big)\\
	-\frac{1}{2}(V_{\bar{X}} + V_X + V_Y), \label{E:Latticepotential}
\end{multline}
where $V_X$, $V_Y$ and $V_{\bar{X}}$ are proportional to the intensities of each of the laser beams \footnote{The expression for the potential used here is the same as Eq. (1) in \cite{Tarruell12} after a $\pi/4$ rotation in the counterclockwise direction and replacing their laser wave vector $k$ by $K=\sqrt{2}k$; after this rotation the lattice vectors become horizontal and vertical. Additionally, since we choose $K$ instead of $k$ to define the recoil energy, our recoil energy is twice the one used in \cite{Tarruell12}.} . We use values of $V_X = 0.25 E_R$, $V_Y = 1.0 E_R$ and $V_{\bar{X}} =3.5 E_R$, where
\begin{equation}
	E_R \equiv \frac{\hbar^2 K^2}{2m} \notag
\end{equation} 
is a recoil energy; the relative phase between lasers associated with $V_X$ and $V_{\bar{X}}$ is given by $\theta$. In addition to $E_R$, another important physical quantity is the recoil velocity
\begin{equation}
	v_R \equiv \frac{\hbar K}{m}, \notag
\end{equation}
which will be used in the velocity plots discussed in \refsecs{S:SemianalyticalResults} and \ref{S:Numerical}. The lattice vectors in real space are in the $\hat{\mbf x}$ and $\hat{\mbf y}$ directions and have magnitude $a = \lambda/\sqrt{2}$, where $\lambda$ is the wavelength of all three lasers; for the experiments described by Tarruell et al. \cite{Tarruell12}, $\lambda = 1064 \text{ nm}$. The reciprocal lattice vectors are in the $\hat{\mbf x}$ and $\hat{\mbf y}$ directions with magnitude 
\begin{equation}
	K = \frac{2\pi}{a}, \notag
\end{equation}
which is also the linear dimension of the Brillouin zone. The potential resembles a squeezed honeycomb lattice and is shown in \reffig{fig:Potential}.\\

%Figure 6
\begin{figure}
\centering
\includegraphics[width=\columnwidth]{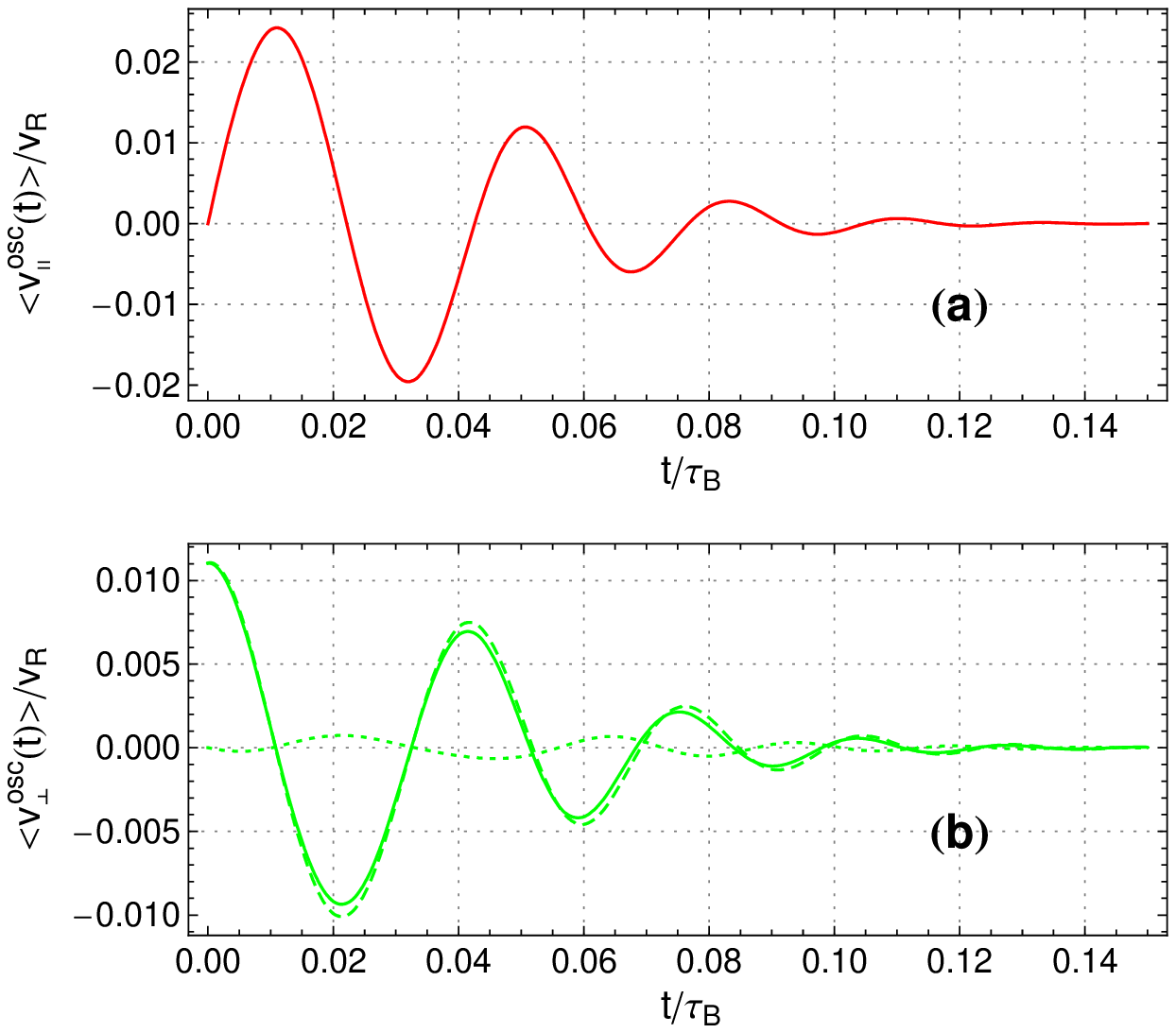}
\caption{(Color online) Decomposition of the oscillating term $\bvoscint{t}$ (solid lines) in group velocity and anomalous velocity contributions, for the example shown in \reffig{fig:DiagPath}; the red (green) curves correspond to the parallel (perpendicular) components of the velocity. In the direction parallel to the force (panel (a)), only the dynamical oscillations from $\mbf V^{\text{g}}_{N}(t)$ contribute. However, in the perpendicular direction (panel (b)) both the dynamical oscillations from $\mbf V^{\text{g}}_{N}(t)$ (dotted line) and the dynamical oscillations from $\mbf V^{\text{an}}_{N}(t)$ (dashed line) contribute; the latter has a more significant contribution in the time range shown here.}
\label{fig:DiagPathDecomp}
\end{figure}

When $\theta$ is set to $\pi$, the lattice satisfies space-inversion symmetry, and the two lowest energy bands intersect each other at two Dirac points. In this case, the local Berry curvature is zero everywhere except at the two Dirac points, where it is singular \cite{Fuchs10}. However, tuning $\theta$ to values slightly different from $\pi$ breaks space-inversion symmetry, opens up a band gap (see \reffig{fig:Band}), and results in a more well-behaved local Berry curvature, (see \reffig{fig:Berry}). It is for this reason that we take $\theta = 1.02\pi$. Although the local Berry curvature is no longer singular, the region of significant local Berry curvature is still very localized. This allows for a great degree of control of the amount of Berry curvature ``seen'' by the wave packet.\\

%************

%*Semianalytical approximation*

\subsection{Semianalytical approximation} \label{S:SemianalyticalResults}

For the envelope function in \refeq{E:InitialWvp} we will use a Gaussian of spread $\sigma = 0.05 K$,
\begin{equation}
	f_N(\mathbf{k}) = \frac{1}{\sqrt{\pi} \sigma} \exp\left(- \frac{(\mathbf{k-k_0})^2}{2\sigma^2} \right),
	\label{eq:gaussian}
\end{equation}
where $\mathbf{k_{0}}$ is the mean of the envelope function. Initially, the wave packet is entirely in the first band, that is $N = 1$. To study the evolution of the wave packet, a force of magnitude $F = \tilde{F} \, K E_R$ will be applied, where $\tilde{F} = 1/2000$ is a dimensionless parameter. With the application of a constant force, the wave packet travels in a straight line through the Brillouin zone.  As described in \reffig{fig:trajectories}, we will consider three trajectories that pass through at least one of the Dirac points where the band gap is the smallest and local Berry curvature the strongest. This allows the wave packet to exhibit more noticeable oscillations associated with the dynamical inverse effective mass tensor and the dynamical anomalous velocity.\\

%Figure 7
\begin{figure}
\centering
\includegraphics[width=\columnwidth]{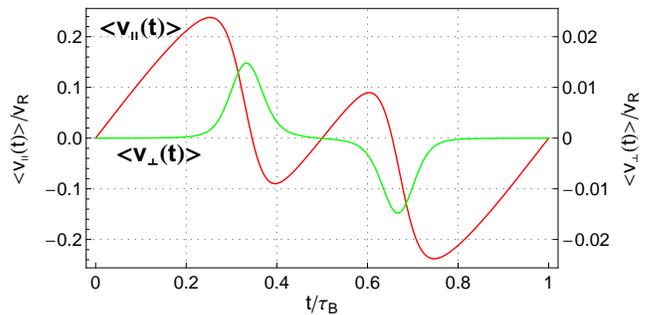}
\caption{(Color online) Expectation value of the velocity from the semianalytical approximation for the same parameters used in \reffig{fig:DiagPath}, but using \textbf{Q} as the starting point for the wave packet (see \reffig{fig:trajectories}); the red (green) curves correspond the parallel (perpendicular) components of the velocity.}
\label{fig:GammaPath}
\end{figure}

It is natural to decompose the full expression for the expectation value of the velocity of the wave packet in the same manner as \refeq{E:VelocityThreeTerms},
\begin{equation}
	\expval{\bvelOp(t)} \approx \bveffint{t} + \bvanint{t} + \bvoscint{t}, \label{E:VelocityDecomp}
\end{equation}
where each of the three terms is an integration of $ \vg{a}{N}{\mbf k} $, $ \van{a}{N}{\mbf k} $, and $ \vosc{a}{N}{\mbf k}{t} $ over the wave packet, respectively. In our first example we start with a wave packet at point \textbf{P} and a force in the direction of path $\alpha$ (see \reffig{fig:trajectories}). The decomposition above is shown in \reffig{fig:DiagPath}. The dashed line corresponds to $\bveffint{t} + \bvanint{t}$, where both the group velocity and the anomalous velocity are taken into account. The solid line gives the full expression for $\expval{\bvelOp(t)}$, with the oscillation term included. As expected from the discussion in \refsec{S:Dynamics}, at $t = 0$ the full expression for the velocity coincides with the prediction by just the group velocity. For $t > 0$, the velocity oscillates around the sum of the group and anomalous velocities. These oscillations decay within a tenth of  the Bloch period, but they reappear near the end of the Bloch oscillation.\\

Because of the high symmetry of path $\alpha$, the group velocity in this example is directed entirely along the direction of the force. This offers the closest analogy to a one-dimensional lattice~\cite{Duque12}. To further explore this analogy, we decompose the dynamical oscillation term $\bvoscint{t}$ in components parallel and orthogonal to the applied force ($\voscintpar{t}$ and $\voscintperp{t}$, respectively) (see \reffig{fig:DiagPathDecomp}).\\

%Figure 8
\begin{figure}
\centering
\includegraphics[width=\columnwidth]{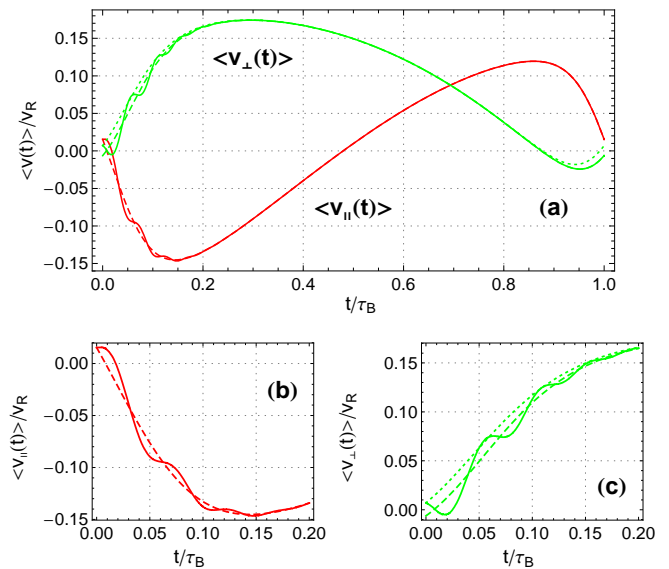}
\caption{(Color online) Expectation value of the velocity from the semianalytical approximation, \refeq{E:VelocityDecomp}, without the oscillating term $\bvoscint{t}$ (dashed lines) and including it (solid lines). In this example the force points in the $\hat{\mbf x}$ direction with $\tilde{F}=1/2000$; the wave packet moves through the Brillouin zone starting at point \textbf{R} and following path $\beta$ (see \reffig{fig:trajectories}). The red curves correspond to $\expval{\text{v}_{\parallel}(t)}$, the component of the velocity parallel to the force, while the green curves correspond to $\expval{\text{v}_{\perp}(t)}$, the component of the velocity perpendicular to the force ($\hat{\mbf y}$ direction). For reference, the component of the group velocity perpendicular to the force is plotted with dotted green lines. (a) Results over one Bloch period, $\tau_B=\sqrt{2} \hbar K / F$. (b-c) Initial behavior of the velocity, showing the oscillations due to $\bvoscint{t}$.}
\label{fig:horizpath}
\end{figure}

The local anomalous velocity is derived from the cross product of the local Berry curvature with the force, and therefore it is orthogonal to the force; similarly, the oscillating term associated with the anomalous velocity is also perpendicular to the force. Thus, all of the contribution to the parallel component of the dynamical oscillation of the velocity comes from oscillations of the group velocity. Interestingly, the converse is not true. Even though for this trajectory the group velocity is strictly parallel to the force, and so the acceleration predicted simply by the inverse effective mass tensor would be in that direction, the dynamical oscillation associated with the group velocity is not confined to this direction for $t>0$. In fact, the dynamical oscillations in both $\mbf V^{\text{g}}(t)$ and $\mbf V^{\text{an}}(t)$ (see \refeqs{E:Vga} and \eqref{E:Vana}) contribute in the direction orthogonal to the force (see \reffig{fig:DiagPathDecomp}). Thus, even for highly symmetric paths there can be oscillations of the velocity perpendicular to the force that are associated with the group velocity, and so the strict analogy to motion in a one-dimensional lattice breaks down.\\

Both the group velocity and anomalous velocity are periodic because they depend only on properties of the band structure, which is periodic. However, the last term of \refeq{E:VelocityDecomp}, associated with $ \vosc{a}{N}{\mbf k}{t}$, is also dependent on the dynamics of the wave packet itself. As such, the full expression for velocity does not exhibit periodicity. This dependence on the dynamics of the wave packet is even more evident when we change the starting point of the trajectory; for example, in \reffig{fig:GammaPath} we choose \textbf{Q} as the starting point but we keep the force in the same direction so that the wave packet moves along the path $\alpha$ (see \reffig{fig:trajectories}). In this case the wave packet ``experiences'' the least amount of Berry curvature at the beginning (see \reffigs{fig:Berry} and \ref{fig:GammaPath}). We see that the group and anomalous velocities are shifted as expected due to the new starting point; the dynamical oscillations in $\mbf V^{\text{g}}(t)$ and $\mbf V^{\text{an}}(t)$, on the other hand, have virtually vanished. From this observation, we conclude that the dynamical oscillations depend on the starting point of the wave packet, even when following the same path.\\

%Figure 9
\begin{figure}
\centering
\includegraphics[width=\columnwidth]{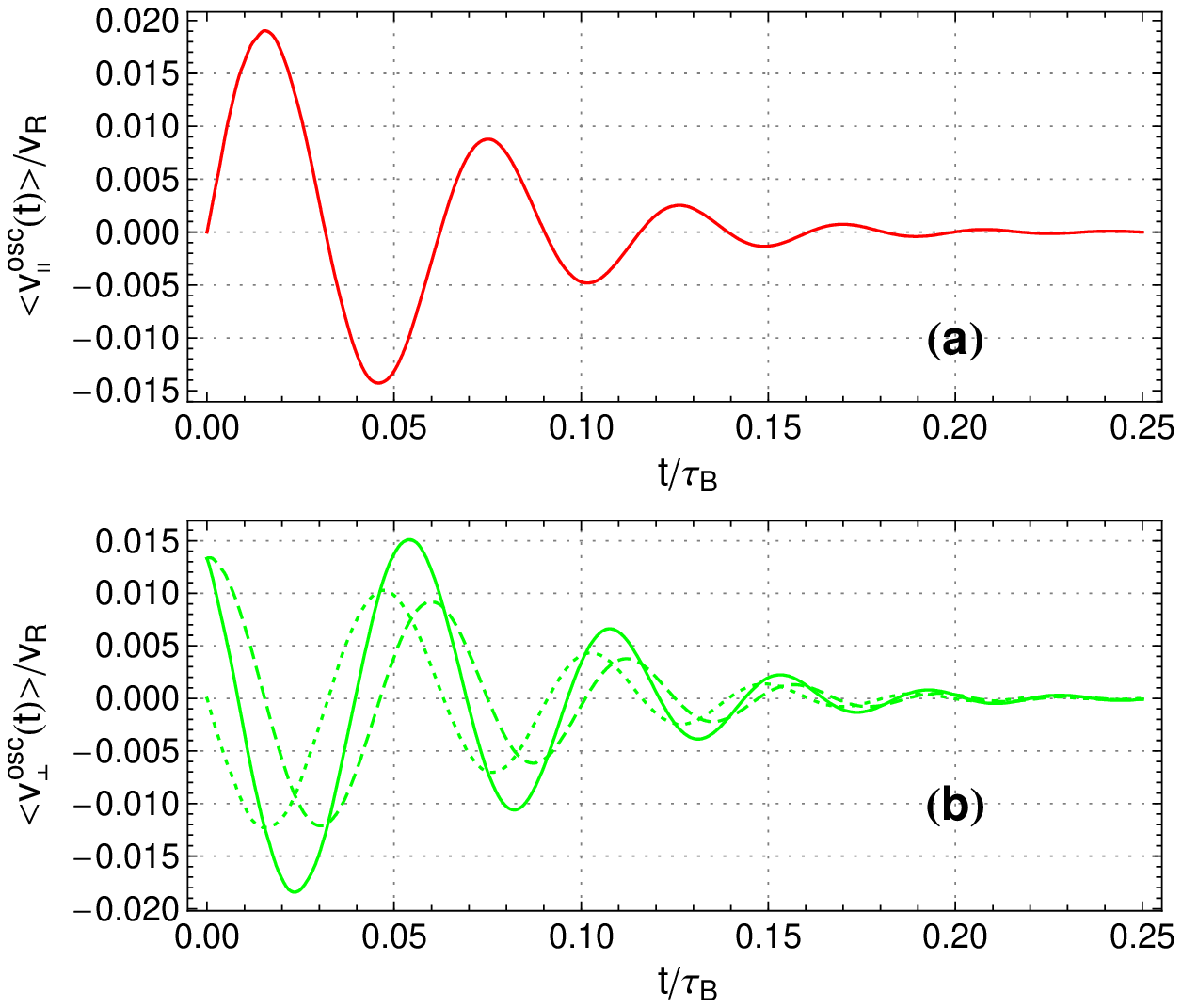}
\caption{(Color online) Decomposition of the oscillating term $\bvoscint{t}$ (solid lines) in its group velocity and anomalous velocity contributions, for the example shown in \reffig{fig:horizpath}; the red (green) curves correspond the parallel (perpendicular) components of the velocity. In the direction parallel to the force (panel (a)), only the dynamical oscillations from $\mbf V^{\text{g}}_{N}(t)$ contribute. However, in the perpendicular direction (panel (b)), both the dynamical oscillations from $\mbf V^{\text{g}}_{N}(t)$ (dotted line) and the dynamical oscillations from $\mbf V^{\text{an}}_{N}(t)$ (dashed line) contribute; in contrast to the behavior in \reffig{fig:DiagPathDecomp}, the two types of oscillations have similar contributions to $\bvoscint{t}$.}
\label{fig:horizpathDecomp}
\end{figure}

For the last example, shown in \reffigs{fig:horizpath} and \ref{fig:horizpathDecomp}, we present a more general path without the symmetries of path $\alpha$. Starting at the point \textbf{R}, we direct the force parallel to path $\beta$ (see \reffig{fig:trajectories}). Unlike in path $\alpha$, here the group velocity does not point exclusively along the direction of the force. The initial behavior shown in \reffig{fig:horizpath} is similar to what was seen in the first example. The velocity starts at the group velocity (dashed line for $\expval{\text{v}_{\parallel}(t)}$ and dotted line for $\expval{\text{v}_{\perp}(t)}$) and oscillates about the combination of the group and anomalous velocities (dashed lines). Despite not being strictly periodic, the dynamical oscillations of the velocity of the wave packet in path $\alpha$ were repetitive, at least qualitatively; we observed dynamical oscillations of similar amplitude and frequency at the start and the end of the Bloch oscillation. This is no longer true in the present example. After the initial dynamical oscillations of the velocity, no more revivals are seen near the end of the first Bloch oscillation; even during the second Bloch period, the dynamical oscillations are completely absent (see \reffig{F:vel_finv2000_pathh}). This lack of revivals can be described as a form of dephasing. The tensor $\tosc{a}{b}{N}{\mbf k}{t}$, responsible for the oscillations, contains a phase $\gamma_{nN}(\bkmov ,t)$ (see \refeq{E:OscillationTensor}). This tensor is integrated over the wave packet as it moves through the Brillouin zone. In general, the phase in $\tosc{a}{b}{N}{\mbf k}{t}$ accumulated by each $\mbf k$-component of the wave packet can be different even after a full Bloch period and we expect to observe dephasing. For the central path $\alpha$, however, the $\mbf k$-components of the wave packet trace pairs of parallel paths that are reflections of each other along the diagonal and acquire the same phase; therefore, with respect to the dephasing, the motion for path $\alpha$ is essentially as in the one-dimensional case, where revivals are observed \cite{Duque12}. This kind of symmetry is not seen for a wave packet moving along a central path $\beta$, which explains why the oscillations decay in this case.\\

%************

%*Comparison with full numerical solution*

\subsection{Comparison with full numerical solution} \label{S:Numerical}

In order to verify the validity of the semianalytical results presented in \refsec{S:SemianalyticalResults}, we compare them with full numerical solutions of the time-dependent Schr\"odinger equation for the Hamiltonian \refeq{E:FullHamiltonian}, with a force suddenly applied at $t=0$ and left constant afterwards. The approximate expression for the wave packet velocity, \refeq{E:VelocityThreeTerms}, is expected to be accurate for forces such that $\foMix_{nN}(\mbf k)$ is small \cite{Adams56, Wannier60, Duque12}; roughly, this requirement means that the energy drop over one unit cell associated with the force should be small compared with the energy gap between the starting band $N$ and its closest neighboring band. For the full numerical calculation we use the split-step operator method \cite{Feit82}. As usual, the kinetic energy term of the Hamiltonian is treated in Fourier space and the potential energy term (including the applied force) is treated in real space, switching back and forth between the two spaces with a Fast Fourier Transform implementation \cite{Frigo05}. The expectation value of the velocity is calculated from the Fourier components of the wave packet. \\

%Figure 10
\begin{figure}
\centering
\includegraphics[width=\columnwidth]{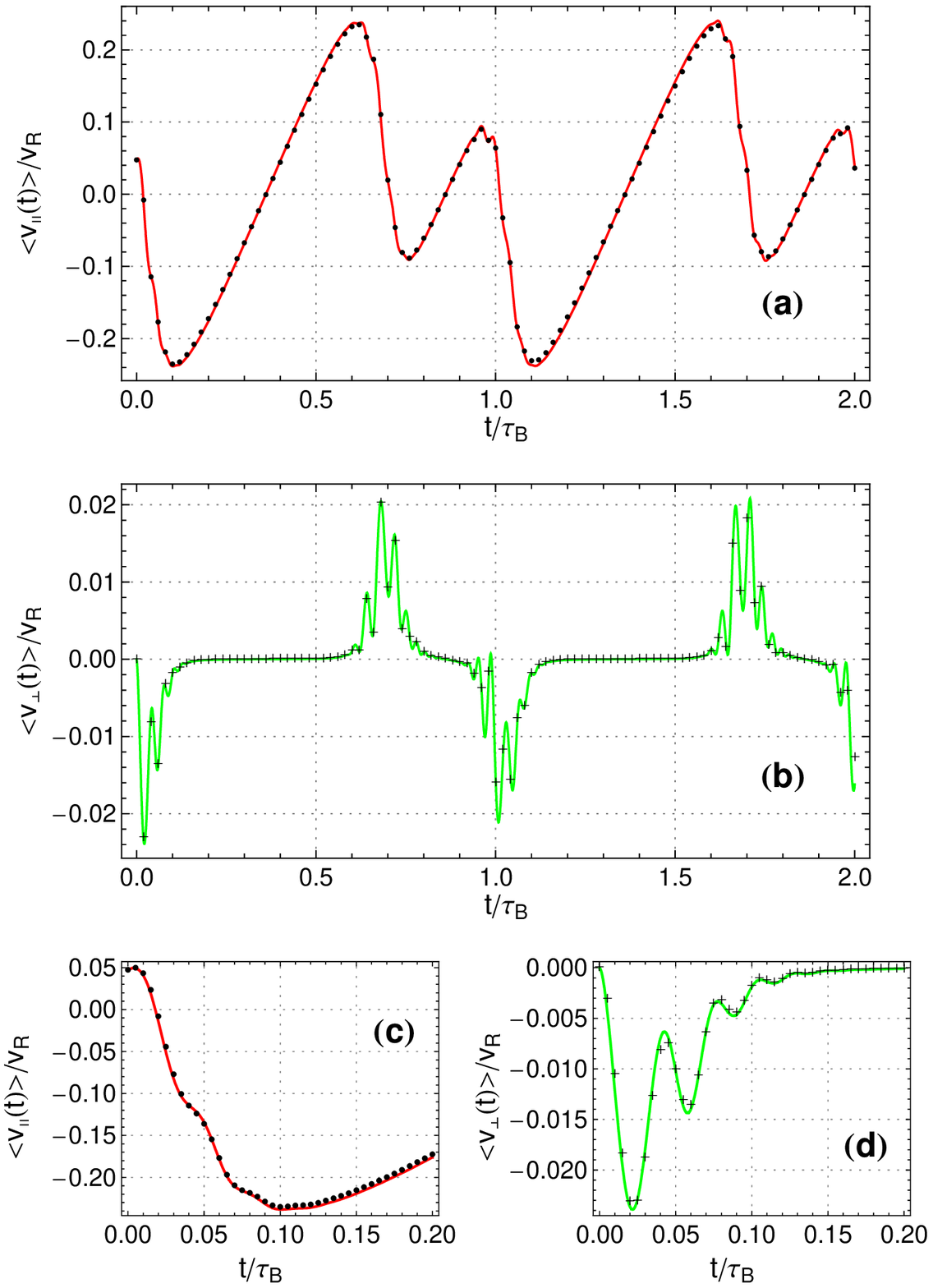}
\caption{(Color online) Comparison between the expectation value of the velocity shown in \reffig{fig:DiagPath} for path $\alpha$ in the Brillouin zone (solid lines) and a full numerical calculation (dots and crosses). The red lines and black dots correspond to the components of the velocity parallel to the force; the green lines and black crosses correspond to the components of the velocity perpendicular to the force. (a-b) Results over two Bloch periods. (c-d) Initial behavior of the velocity, showing the oscillations due to $\bvoscint{t}$ (see \refeq{E:VelocityDecomp}).}
\label{F:vel_finv2000_pathdd}
\end{figure}

\begin{figure}
\centering
\includegraphics[width= \columnwidth]{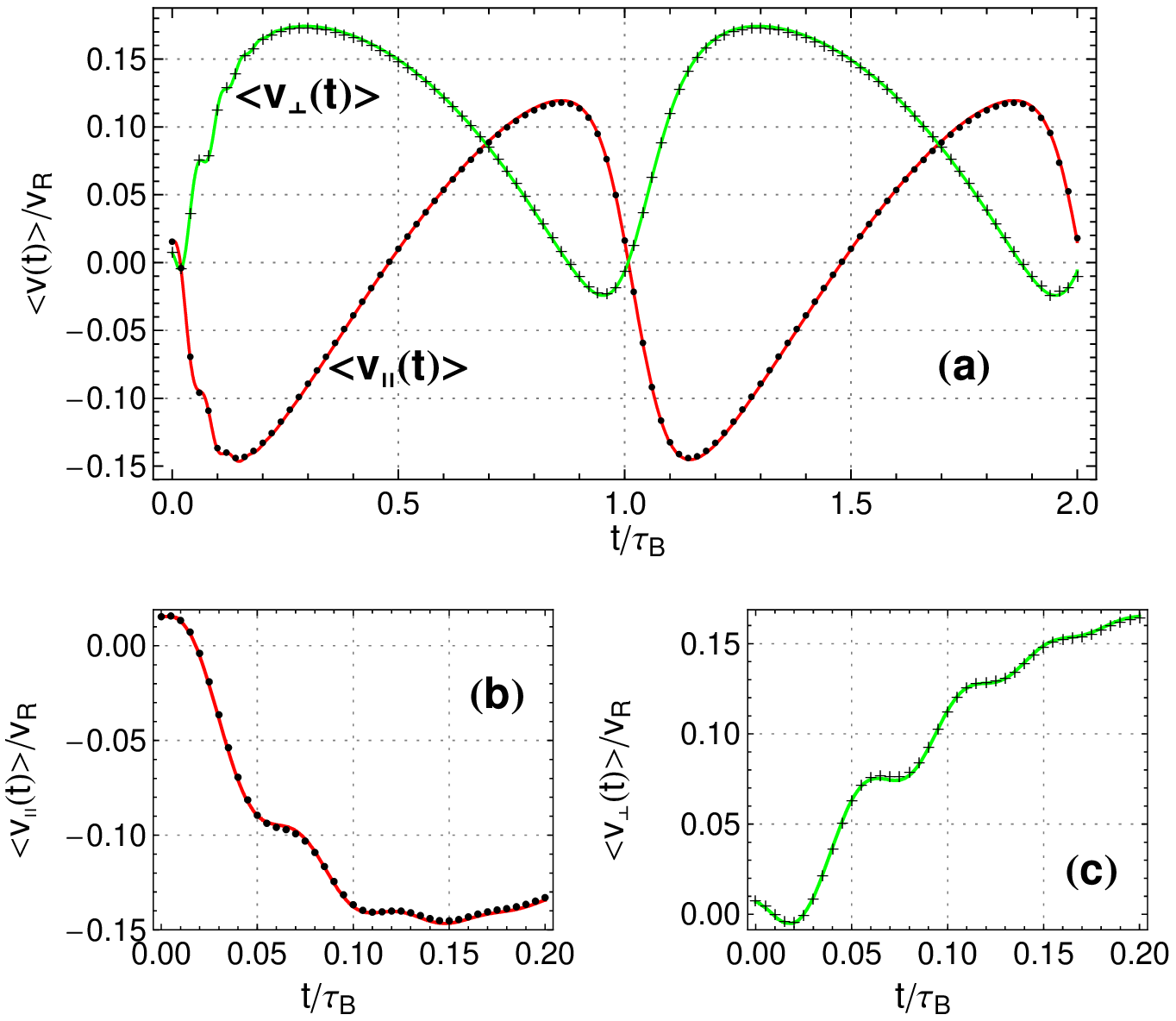}
\caption{(Color online) Comparison between the expectation value of the velocity shown in \reffig{fig:horizpath} for path $\beta$ in the Brillouin zone (solid lines) and a full numerical calculation (dots and crosses). The red lines and black dots correspond to the components of the velocity parallel to the force; the green lines and black crosses correspond to the components of the velocity perpendicular to the force. (a) Results over two Bloch periods. (b-c) Initial behavior of the velocity, showing more clearly the oscillations due to $\bvoscint{t}$ (see \refeq{E:VelocityDecomp}).}
\label{F:vel_finv2000_pathh}
\end{figure}

In \reffigs{F:vel_finv2000_pathdd} and \ref{F:vel_finv2000_pathh} we compare the expectation value of the velocity for paths $\alpha$ and $\beta$ in the Brillouin zone presented in \reffigs{fig:DiagPath} and \ref{fig:horizpath} with full numerical calculations over two Bloch periods. Note the excellent agreement between the two approaches, over short and long time scales. The force used in these examples is small enough to guarantee that the semianalytical expression  \refeq{E:VelocityThreeTerms} predicts correctly the oscillations associated with the dynamics of the effective mass and anomalous trasport. Furthermore, the presence of revivals for path $\alpha$ (see \reffig{F:vel_finv2000_pathdd}) and their absence for path $\beta$ (see \reffig{F:vel_finv2000_pathh}) is confirmed by the full numerical calculation.\\

%Figure 11

\begin{figure*}
\centering
\includegraphics[width=16cm]{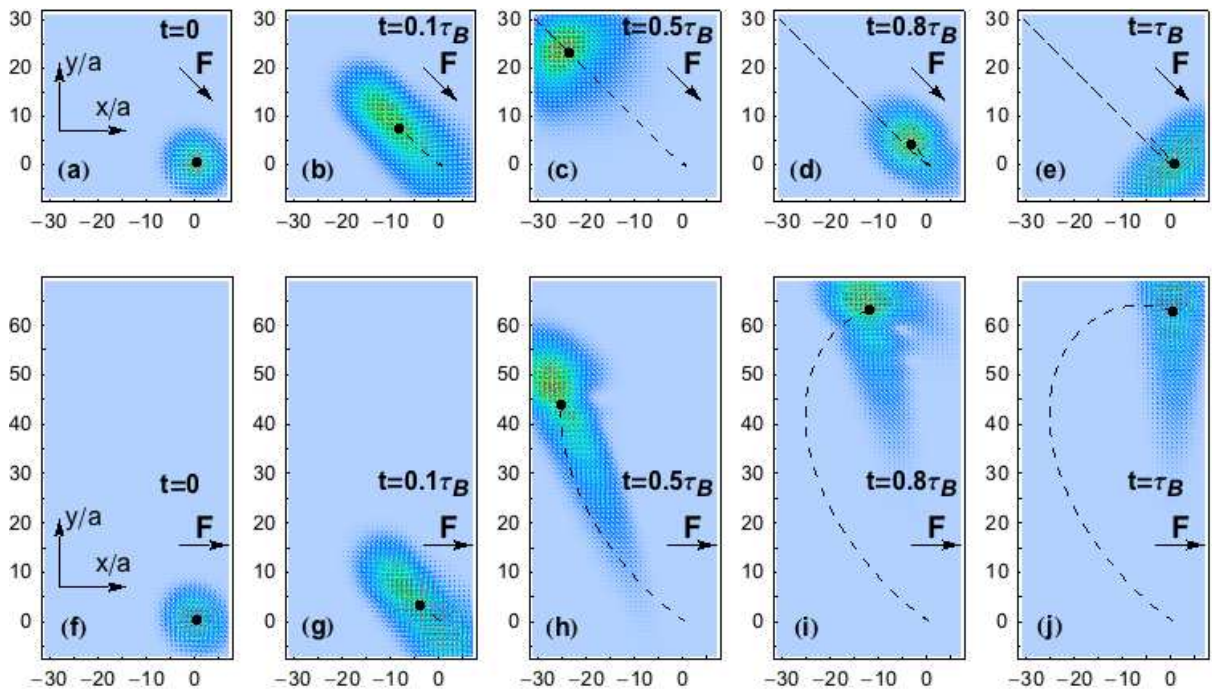}
\caption{(Color online) Time evolution of the absolute value of the wave packet in real space calculated using the split-step operator method for (a-e) path $\alpha$ and (f-j) path $\beta$ in the Brillouin zone. The parameters are the same as the ones used in \reffigs{F:vel_finv2000_pathdd} and \ref{F:vel_finv2000_pathh}. The black dots mark the expectation value of the position for the snapshot time, and the dashed line shows the trajectory starting at $t=0$. The horizontal and vertical axes show position in units of the lattice constant.}
\label{F:wvp_ev}
\end{figure*}

We can also illustrate the evolution of the wave packet in real space using the results from the time propagation with the split-step operator method. In \reffig{F:wvp_ev} we show snapshots of the wave packet in the two situations considered in \reffigs{F:vel_finv2000_pathdd} and \ref{F:vel_finv2000_pathh}, for different times within one Bloch period. In both cases the initial wave packets are the same, but in time the trajectory (dashed line in \reffig{F:wvp_ev}) and dispersion of the wave packets evolve differently. Due to the symmetry of path $\alpha$ (see \reffig{fig:trajectories}), the group velocity is always parallel or antiparallel to the force, resulting in an oscillation along that direction (see \reffigs{F:wvp_ev}a-e). On the other hand, for path $\beta$, the wave packet traces a more complicated trajectory in real space, and it does not return to its starting point (see \reffigs{F:wvp_ev}f-j). The anomalous velocity and the dynamical corrections modify the real space trajectories described by the group velocity alone, but this change is small compared with the scale of the trajectories shown in \reffig{F:wvp_ev} over one Bloch period. With respect to the dispersion of the wave packet, the spread is more pronounced for path $\beta$ than path $\alpha$ (compare \reffigs{F:wvp_ev}c and \reffig{F:wvp_ev}h), and the shape of the wave packet remains more symmetric around its center for path $\alpha$ (compare \reffigs{F:wvp_ev}c-e and \reffigs{F:wvp_ev}h-j).\\

%Figure 12

\begin{figure}
\centering
\includegraphics[width= \columnwidth]{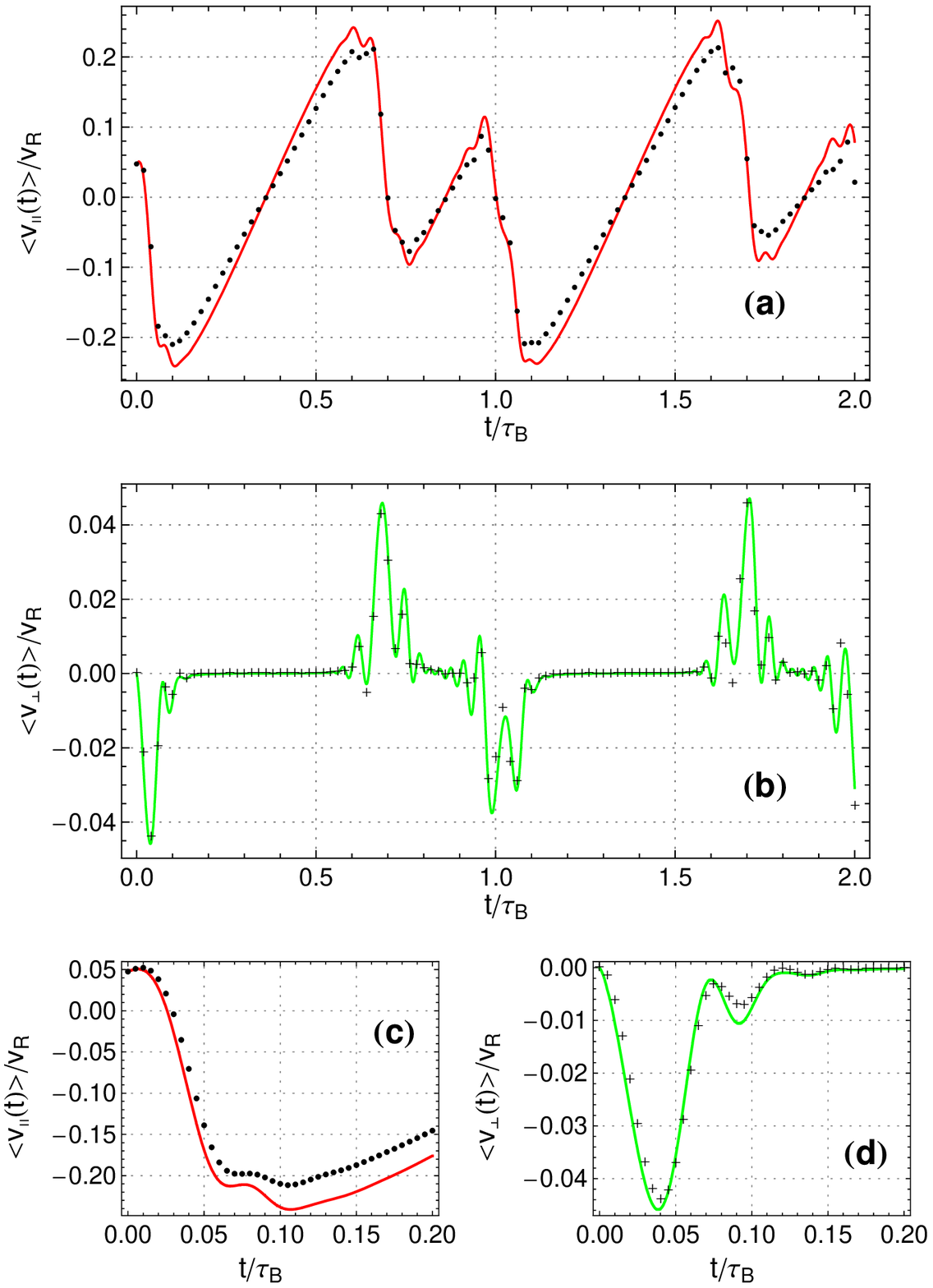}
\caption{(Color online) Comparison between the semianalytical approximation (solid lines) and the full numerical calculation (dots and crosses) for the same parameters as in \reffig{fig:DiagPath} but doubling the force ($\tilde{F} = 1/1000$). The red lines and black dots correspond to the components of the velocity parallel to the force; the green lines and black crosses correspond to the components of the velocity perpendicular to the force. (a-b) Results over two Bloch periods. (c-d) Initial behavior of the velocity, showing the oscillations due to $\bvoscint{t}$ (see \refeq{E:VelocityDecomp}).}
\label{F:vel_finv1000_pathdd}
\end{figure}

The ultimate breakdown of the semianalytical expression \refeq{E:VelocityThreeTerms} is associated with Zener tunneling, the probability of which increases for larger forces. Wannier's method of decoupling the bands in the presence of an applied force \cite{Wannier60} cannot describe Zener tunneling even if higher orders of his power expansion are considered \cite{Nenciu91}. Accordingly, the picture presented in \refsec{S:TheoreticalFramework} is only valid for wave packets mainly in one band with small amplitudes over neighboring bands, which is the typical requirement in the semiclassical description of transport. The start of the breakdown of the semianalytical approximation is shown in \reffig{F:vel_finv1000_pathdd} for path $\alpha$ (see \reffig{fig:trajectories}) and a force twice as large as the one used so far. In this case the most significant difference between the semianalytical approximation and the full numerical calculation appears in the component of the velocity parallel to the force, as the semianalytical result overestimates the amplitude of the Bloch oscillation of the usual group velocity. This deviation occurs early in the evolution of the expectation value of the velocity, since the starting point of the trajectory in the Brillouin zone is near one of the Dirac points, where the first two bands are close and Zener tunneling is more probable. Nevertheless, note that \refeq{E:VelocityThreeTerms} is still a good approximation, as it agrees at least qualitatively with the full numerical calculation.\\

%Figure 13

\begin{figure}
\centering
\includegraphics[width=8cm]{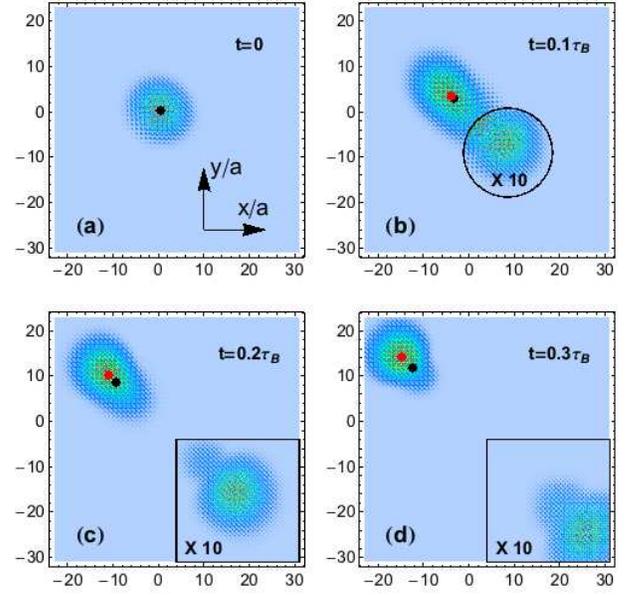}
\caption{(Color online) Snapshots of the absolute value of the wave packet for the parameters used in \reffig{F:vel_finv1000_pathdd} calculated from the time evolution using the split-step operator method. At $t=0.1 \TauB$ a wavelet starts to form and afterwards it moves towards the lower right corner of the real space window. The black and red dots mark the expectation value of the position for the snapshot time, calculated from the full numerical and semianalytical calculations, respectively. In (b-d) the smaller frames show the absolute value of the wave packet amplified ten times. The horizontal and vertical axes show position in units of the lattice constant.}
\label{F:zener_wvp}
\end{figure}

In the semianalytical approximation the wave packet $\ketPsi{t}$ is a superposition of a main MBS wave packet associated with band $N$ (see \refeq{E:AmplitudeMain}) and MBS wave packets with smaller amplitudes associated with neighboring bands, $n \ne N$ (see \refeq{E:AmplitudeNeighbor}). This suggests that the initial wave packet in real space will split into a main wave packet associated with band $N$ and small ones associated with $n \ne N$; these wave packets will move differently according to the properties of the band to which they correspond. The presence of this splitting is confirmed by the full numerical calculation, as illustrated in \reffig{F:zener_wvp} for the parameters used in \reffig{F:vel_finv1000_pathdd}. The snapshots in \reffig{F:zener_wvp} show a main wave packet associated with band $N=1$ and a small wave packet associated with the next band, $n=2$, which moves in the opposite direction to the main wave packet.\\

%Figure 14

\begin{figure}
\centering
\includegraphics[width=8cm]{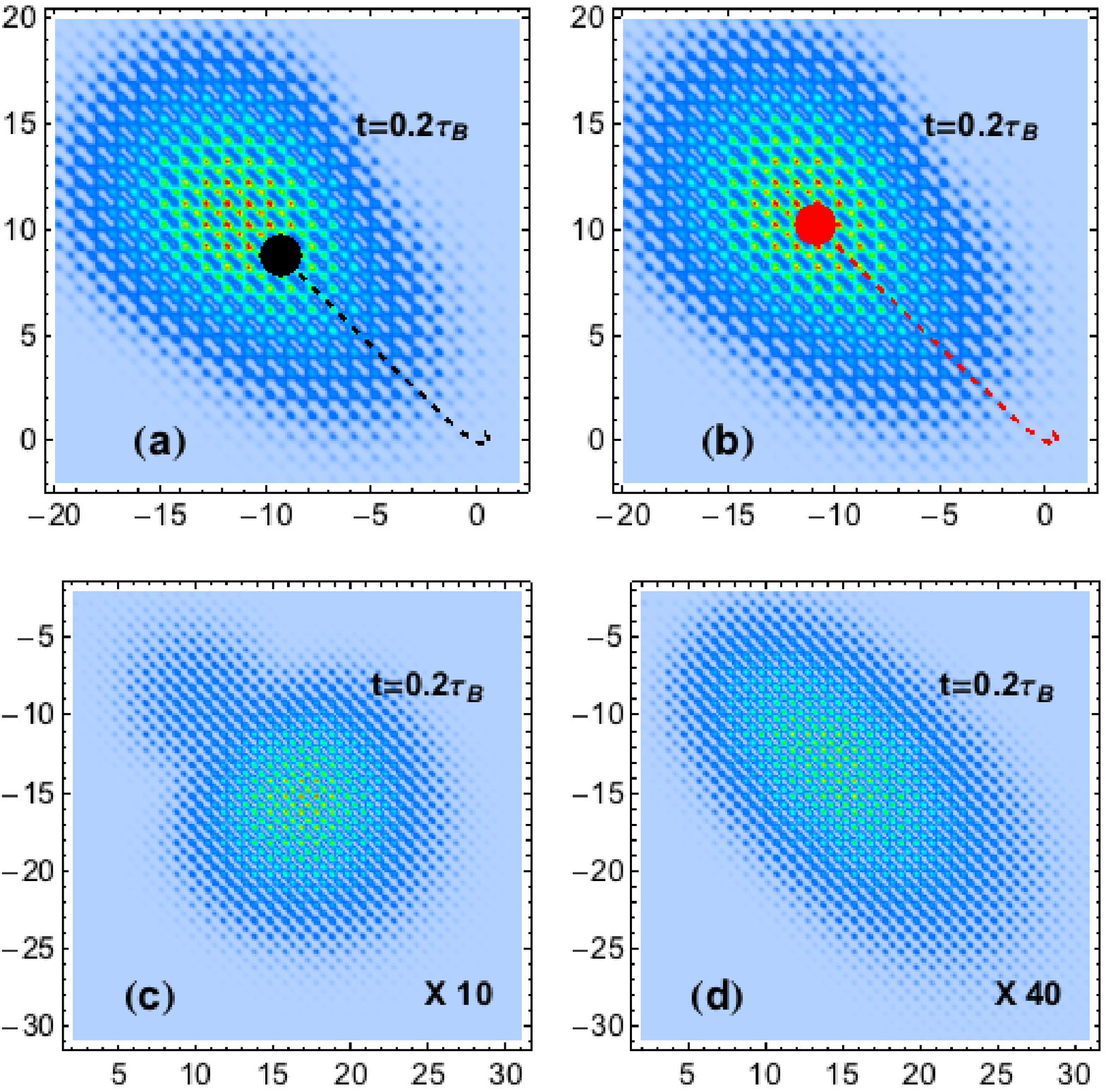}
\caption{(Color online) Detailed view of the wave packet shown in \reffig{F:zener_wvp}c. The black and red dots (and the dashed lines) mark the expectation value of the position calculated from the full numerical and semianalytical calculations, respectively. The horizontal and vertical axes show position in units of the lattice constant. (a-b) Main wave packet from the full numerical and semianalytical calculations, respectively. (c-d) Small wave packet from the full numerical and semianalytical calculations, respectively.}
\label{F:comparison_wvp}
\end{figure}

Even thought the semianalytical approximation predicts the splitting of the initial wave packet, it cannot describe correctly the amplitude and shape of the small wave packet for strong forces, such as the one used in \reffig{F:zener_wvp}. The failure of the semianalytical approach in this example is shown in \reffig{F:comparison_wvp}, where we compare the wave packets calculated using this approximation and the results from the full numerical calculation. Note that the main wave packet is essentially the same in the two approaches (compare \reffigs{F:comparison_wvp}a and b), but the semianalytical result predicts a small wave packet with a different shape and underestimates its amplitude (compare \reffigs{F:comparison_wvp}c and d). Consequently, the expectation value of the position differs in the two calculations as can be seen in \reffigs{F:zener_wvp} and \ref{F:comparison_wvp}, where the expectation value of the position calculated with the semianalytical approximation (red dots) is shifted in the direction of the small wave packet for the full numerical calculation (black dots). The incorrect description of the small wave packet by the semianalytical approximation is responsible for the overestimation of the group velocity calculated with this method in \reffig{F:vel_finv1000_pathdd}. As the applied force is increased, Zener tunneling becomes more important and the amplitudes of the wave packets associated with the neighboring bands, $n \ne N$, increase; consequently, the weight of these amplitudes modifies more significantly the expectation values of position and velocity predicted by the main wave packet alone. Since the semianalytical approximation cannot predict correctly the contribution of these wave packets, the dynamics calculated with this approach become less accurate.\\

%Figure 15

%************************

%**CONCLUSION**

\section{CONCLUSION} \label{S:Conclusion}

We have discussed the dynamics of a wave packet in a periodic potential prepared in one band and subject to the sudden application of a uniform force, which remains constant afterwards. We have found that the usual semiclassical description, involving the inverse effective mass tensor and the anomalous velocity, requires corrections. When the force is suddenly applied, the particle responds initially as if it were free; its acceleration is characterized by the bare mass (see \refeq{E:InitialInvEffMassTensor}), and there is no anomalous velocity (see \refeq{E:InitialGroupAnomVel}). However, it is possible to define dynamical quantities associated with the inverse effective mass tensor (see \refeq{E:Mstar}) and the anomalous velocity (see \refeq{E:DynAnomVel}). These quantities initially take the values that would characterize a free particle; at later times they oscillate about the usual expressions for these quantities as the wave packet moves through the Brillouin zone. The total velocity of the wave packet, \refeq{E:vuse}, includes a dynamical group velocity, associated with the dynamical inverse effective mass tensor, and the aforementioned dynamical anomalous velocity. Even for cases when the usual inverse effective mass tensor predicts an acceleration parallel to the applied force (for example, in the path $\alpha$ shown in \reffig{fig:trajectories}), the dynamical inverse effective mass tensor allows for oscillations of the velocity parallel and perpendicular to the force (see \reffig{fig:DiagPathDecomp}). In addition to the acceleration described by the dynamical inverse effective mass tensor, there is a dynamical anomalous acceleration associated with the dynamical anomalous velocity (see \refeq{E:DecompAcceleration}); both these dynamical anomalous quantities are always perpendicular to the applied force.\\

We have derived semianalytic expressions for all these dynamical quantities, and calculated them for a particle subject to a suddenly applied force in a two-dimensional optical lattice \cite{Tarruell12}. Besides exhibiting aspects of wave packet motion involving the topology of the bands, which do not arise in one-dimensional lattices, the wave packet motion in the two-dimensional lattice shows interesting features in the interplay between the Bloch oscillations and the dynamics of the group and anomalous velocities. In one-dimensional lattices it was shown that the initial dynamical oscillations have revivals after a Bloch period as a result of the cyclic path of the wave packet in the Brillouin zone \cite{Duque12}. In the two-dimensional lattice considered here, we showed that not every cyclic path in the Brillouin zone leads to revivals after one Bloch period (see \reffig{F:vel_finv2000_pathh}); these revivals only occur for paths where the symmetry of the band structure allows each $\mbf{k}$-component of the wave packet to accumulate similar phases over a Bloch period (see \reffig{F:vel_finv2000_pathdd}). This behavior shows that, while the group velocity and the Berry curvature have a periodicity given by the Bloch period, the dynamical oscillations discussed here do not display such periodicity. Revivals are still possible in the two-dimensional lattice for some paths, but the dynamical oscillations are still not periodic over one Bloch period. Furthermore, these oscillations depend on the starting location on the chosen path in the Brillouin zone (compare \reffigs{fig:DiagPath} and \ref{fig:GammaPath}).\\

The results from the semianalytical approximation were confirmed by a full numerical solution of the dynamics of the wave packet. The agreement breaks down for strong forces, due to the limitations of the modified Bloch states to decouple completely the bands in the presence of an applied force \cite{Nenciu91}.  In real space the wave packet splits into a main wave packet, associated with the original initial band, and a small wave packet, associated with the next neighboring band (see \reffig{F:zener_wvp}); the semianalytical approximation fails to capture correctly the amplitude and shape of the small wave packet affecting the expectation values of position and velocity calculated with this method (see \reffig{F:comparison_wvp}). However, since the main wave packet is well described by the semianalytical approximation, we find that even for a strong force the prediction of the dynamical oscillations given by this approximation is at least qualitatively correct (see \reffig{F:vel_finv1000_pathdd}).\\

Two-dimensional optical lattices are readily available, suggesting that the dynamics described here can be observed experimentally in this type of system; this would generalize and extend the recent experimental study of these dynamics in a one dimensional optical lattice \cite{Rockson14}. For the application of a force to be ``sudden" requires in practice that the time scale for its appearance is short compared to the time associated with the energy difference between the band of the initial wave packet and the nearest neighboring band. In solid-state systems the time scales and the difficulties in controlling the properties of the lattice have prohibited the observation of the dynamical inverse effective mass tensor, even though some deviations from the usual effective mass behavior have been attributed to its dynamical oscillations \cite{Zhu08}; nonetheless, attosecond science is pushing the timescales on which carrier dynamics in solids can be observed to the sub-femtosecond regime \cite{Ghimire11, Isanov13, Schubert14}. We believe that these developments, combined with the growing interest in topological properties of periodic potentials and their dynamical consequences \cite{Niu10, Virk11, Murakawa2013, Yang13}, make the oscillations discussed here an interesting phenomenon to be studied experimentally both in optical lattices and in solid-state systems.

%\bibliography{mybib}{}


\end{document}